\documentclass[
 reprint,
 superscriptaddress,
 groupedaddress,
preprintnumbers,
nofootinbib,
 amsmath,amssymb,
 aps,
floatfix,
]{revtex4-2}

\usepackage{color, colortbl}
\definecolor{Gray}{gray}{0.9}
\usepackage{siunitx}
\usepackage{makecell}
\usepackage{multirow}

\usepackage{hyperref}
\hypersetup{
  colorlinks   = true,    
  urlcolor     = blue,    
  linkcolor    = blue,    
  citecolor    = red      
}
\usepackage{graphicx}
\usepackage{dcolumn}
\usepackage{bm}

\usepackage{xcolor}

\def \MMn {{\rm MM}_{\rm nbhd}}
\def \MM {{\rm MM}}
\def \a {\alpha}
\def \b {\beta}

\def \no {\nonumber}

\def \Nsg {N_{\rm seg}}
\def \tsg {t_{\rm seg}}
\def \ffl {f_{\rm flat}}
\def \fc {f_{\rm coarse}}
\def \Nt {N_{\rm temp}^{\rm flat}}
\def \Ns {N_{\rm temp}^{\rm stage1}}
\def \Nss {N_{\rm temp}^{\rm stage2}}
\def \Ofl {O_{\rm flat}}
\def \Oc {O_{\rm coarse}}
\def \M {\mathcal{M}}

\def\be{\begin{equation}}
\def\ee{\end{equation}}
\def\bea{\begin{eqnarray}}
\def\eea{\end{eqnarray}}

\begin{document}

\preprint{APS/123-QED}

\title{Hierarchical search for compact binary coalescences in the Advanced LIGO's first two observing runs}

\author{Kanchan Soni}
\email{kanchansoni@iucaa.in}
\affiliation{
Inter-University Centre for Astronomy and Astrophysics, Pune 411007, India} 
 
\author{Bhooshan Uday Gadre}%
\email{bhooshan.gadre@aei.mpg.de}
\affiliation{%
Max Planck Institute for Gravitational Physics (Albert Einstein Institute), D-14476 Potsdam, Germany
}%

\author{Sanjit Mitra}
\email{sanjit@iucaa.in}
\affiliation{
Inter-University Centre for Astronomy and Astrophysics, Pune 411007, India
}%

\author{Sanjeev Dhurandhar}
\email{sanjeev@iucaa.in}
\affiliation{%
Inter-University Centre for Astronomy and Astrophysics, Pune 411007, India
}

\begin{abstract}

Detection of many compact binary coalescences (CBCs) is one of the primary goals of the present and future ground-based gravitational-wave (GW) detectors. While increasing the detectors' sensitivities will be crucial in achieving this, efficient data analysis strategies can play a vital role. With given computational power in hand, efficient data analysis techniques can expand the size and dimensionality of the parameter space to search for a variety of GW sources. Matched filtering-based analyses that depend on modeled signals to produce adequate signal-to-noise ratios for signal detection may miss them if the parameter space is too restrained. Specifically, the CBC search is currently limited to nonprecessing binaries only, where the spins of the components are either aligned or antialigned to the orbital angular momentum. A hierarchical search for CBCs is thus well motivated. The first stage of this search is performed by matched filtering coarsely sampled data with a coarse template bank to look for candidate events. These candidates are then followed up for a finer search around the vicinity of an event's parameter space found in the first stage. Performing such a search leads to enormous savings in the computational cost without much loss in sensitivity. Here we report the first successful implementation of the hierarchical search as a PyCBC-based production pipeline to perform a complete analysis of Laser Interferometer Gravitational Wave Observatory (LIGO) observing runs. With this, we analyze Advanced LIGO's first and second observing run data. We recover all the events detected by the PyCBC (flat) search in the first GW catalog, GWTC-1, published by the LIGO-Virgo collaboration, with nearly the same significance using a scaled background.  In the analysis, we get an impressive factor of 20 reduction in computation compared to the flat search. With a standard injection study, we show that the sensitivity of the hierarchical search remains comparable to the flat search within the error bars. 

\end{abstract}

\maketitle

\section{\label{sec:level1}Introduction\protect}
Gravitational-wave (GW) astronomy began with the detection of GW signal from a binary black hole (BBH) merger, GW150914~\cite{firstevent}, using the Advanced Laser Interferometer Gravitational Wave Observatory (LIGO)~\cite{advligo1,advligo2} observatories. With the latest advancements in the sensitivities of detectors and search techniques like cWB~\cite{cwb}, GstLAL~\cite{gstlal1}, PyCBC \cite{usman}, LIGO-Virgo (LV) collaboration detected GW signals from ten BBHs and one binary neutron star (BNS) coalescence in the first two, O1 and O2, observing runs~\cite{gwtc1}. During this period, several independent searches~\cite{ogc1,ogc2,venumadhav1} over publicly available data detected a few additional BBH events. A paradigm shift in the number of detections occurred with the improvement in the sensitivities of Advanced LIGO~\cite{asi} and Advanced Virgo~\cite{advancedvirgo} detectors in the third observing run. This has led to the detections of many GW events~\cite{gwtc2,gwtc2.1,gwtc3} including GW190425~\cite{GW190425}, the second BNS event, GW190412~\cite{GW190412}, and GW190814~\cite{GW190814}, the first two highly asymmetric compact binary coalescences (CBCs) that emit a significant amount of gravitational radiation beyond the quadrupole moment, and GW190521~\cite{IMBH}, the first binary merger to form an intermediate-mass black hole. 

Matched filtering~\cite{svd,svd-satya,svd-schutz,sathya-owen,findchirp}, a primary and most sensitive algorithm, is used to detect signals that can be well modeled. Since the GW signals from merging binaries in circular orbits can be modeled using their intrinsic\footnote{Component masses ($m_1,m_2$) and individual spins ($\vec{s}_1,\vec{s}_2$) vectors of the coalescing binary.} and extrinsic\footnote{Sky location ($\zeta,\phi$), luminosity distance ($d_L$), orbital inclination ($\iota$), polarization angle ($\psi$), and time and phase of coalescence ($t_c,\phi_c$) of the coalescing binary with respect to the detector.} parameters, the matched-filtering method is employed for their detection. The method involves correlating an interferometer's output, time-series data, with the modeled waveforms (\textit{templates}) for each detector in the network. If a GW signal is present in the detector's output, the correlation results in a peak (\textit{trigger}) in the signal-to-noise ratio (SNR) corresponding to the best-matching template. Since the prior knowledge of the source parameters, like its component masses, spins, and location in the sky, remains unknown to the observers, the search is required to be performed over a wide range of several source parameters using a ``bank of templates." The templates in the bank are closely placed to ensure that the search does not miss any signal. Since the data contain non-Gaussian noise, a coincidence search over the time of arrival, phase, and other source parameters is performed between different detectors to reduce the rate of false alarms. The coincident candidates obtained are then assigned significance based on the noise background.

The above procedure for detecting GW signals from CBCs is followed by the search pipelines like GstLAL~\cite{gstlal1,gstlal2,gstlal3}, MBTA~\cite{mbta1,mbta2}, PyCBC~\cite{usman,gareth,ogc1,ogc2}, and SPIIR~\cite{spiir}. These pipelines perform a \textit{one-step search}\footnote{Search involving match-filtering data, sampled at a fixed rate using a bank of templates} for the nonprecessing coalescing binaries in quasicircular orbits.
 

One of the challenges that template-based search pipelines face is the high computational cost of matched filtering, typically a year's worth of data over $\sim\mathcal{O}(10^5)$ templates. Since this process, especially in the PyCBC (or flat) search, involves fast Fourier Transform (FFT) of the product of uniformly sampled time-series data and a template, the number of floating-point operations scales as $N  \log_2 N$, where $N$ is the number of data points. These operations repeat over $\sim\mathcal{O}(10^5)$ templates~\cite{tito-ian_templatebank}, even in the restricted parameter space of nonprecessing binaries with quasicircular orbits, which amounts to sizable computational cost. The cost further increases when a search is envisaged for precessing binaries where the orbital plane precesses due to the misalignment of component spins with the orbital angular momentum. In such cases, the number of templates and the matched-filtering operations increases at least tenfold~\cite{ianharry}, thus making the search computationally expensive to pursue with the current capabilities. Furthermore, the search for primordial black holes in the subsolar region requires templates $\sim\mathcal{O}(10^5-10^6)$~\cite{subsolar_lvk19,subsolar_lvk21}, which makes the search more expensive. To reduce the search's cost, matched filtering over the data is performed above a frequency of 45 Hz~\cite{subsolar_lvk19,subsolar_lvk21} while compromising with the overall reduction of $\sim24\%$ in the sensitive volume. While another~\cite{nitz_wang_subsolar} search still uses a lower frequency of 20 Hz with waveforms having low eccentricities, it assumes very low nonprecessing spins to make searches computationally manageable. These limitations can be reduced if faster matched-filtering search algorithms are developed.

With the advancements in current detectors and upcoming new detectors, e.g., KAGRA~\cite{kagra} and LIGO-India~\cite{ligoindia}, the CBC detection rate is bound to increase, and finer details of the detected sources would be sought to unravel their exact dynamics, formation, and evolution scenarios. However, this would significantly increase the volume of the search parameter space. The increment in volume would happen in two ways; first, the number of parameters (dimension of the parameter space) of different CBC sources would increase, and secondly, their ranges may increase. Nevertheless, a comprehensive matched-filter based search is important because one would like to capture the nontrivial dynamics of interesting astrophysical sources. Therefore, to facilitate this quest, we assert that a matched-filter based search pipeline needs speeding up by orders of magnitude. 

One way to speed up the search is by performing matched filtering \textit{hierarchically using multiple banks of varying densities}. The idea of performing matched filtering in hierarchical steps was formally introduced in~\citet{mohanty96}, where hierarchy was performed over the chirp mass of binaries using Newtonian waveforms. This work was later extended to the post-Newtonian waveforms~\cite{mohanty98}, where hierarchy was performed over the component masses of a binary system. A further improvement was realized by reducing the sampling rate in the first stage of the hierarchy. In the recent work of~\citet{bug} hierarchy was performed over all the three intrinsic parameters, including the effective spin of the binary. This algorithm had used two-detector coincidence analysis and had provided an order of magnitude speed-up compared to the flat analysis.  

In this paper, we revisit the hierarchical search formulated in~\citet{bug}, and for {\it the first time, implement it as a working PyCBC-based pipeline} to analyze the data from an entire LIGO observing run. We describe an efficient two-stage hierarchical search pipeline to search for GW signals from CBCs in the two detectors. This pipeline improves the hierarchical search sensitivity by incorporating better detection statistics for the single-detector and coincident triggers, as used by the flat analysis in~\citet{gwtc2}. For this work, we construct two template banks--- $coarse$ and $neighborhood~(nbhd)$ bank, to target GW signals from nonprecessing CBC sources that have quasicircular orbits. Using the former bank in the first stage and a dynamical subset of the latter bank in the second stage of the hierarchical search, we test the potency of the pipeline by applying it to the data from the first two observing runs of Advanced LIGO. Our pipeline recovers all the GW events observed by the flat search from the first gravitational wave catalog (GWTC-1)~\cite{gwtc1}. 

In our work, we assign the significance to the detected events using a \textit{scaled background}~\cite{bug}, constructed by scaling the background obtained in the first stage by time sliding the filtered output across detectors using a coarse bank. To justify the accuracy of this background, we perform simulations that involve the injection of the GW signals into the data and compare their recoveries with the hierarchical and flat search separately. Furthermore, we compare the sensitivities of the two searches through ``volume-time" ($VT$) ratio curves. Lastly, we conclude our findings from the injection study by discussing the two searches' matched-filter computational costs.

The paper is organized as follows: 
\begin{itemize}
    \item In Sec. \ref{sec:preq} we state the prerequisites and describe the search methodology for the hierarchical search. The section segregates into subparts. In Sec. \ref{subsec:templatebank}, we describe the generation of template banks. Section \ref{sec:matchedfilter} elaborates on the matched-filtering process and selection criteria for the generated triggers in two stages. The strategy to collect coincident triggers is described in Sec. \ref{sec:ranking}. The final step in the pipeline is to assign significance to the coincident candidates. We describe this process in Sec. \ref{sec:significance}.
    \item In Sec. \ref{sec:pipeline}, we implement the hierarchical search pipeline over the first two observing runs of Advanced LIGO and present our findings. 
    \item In Sec. \ref{sec:comparison},  we compare the sensitivities of the hierarchical search with flat search. We also discuss the computational advantages of the former search with the latter. 
    \item In Sec. \ref{sec:conclusion}, we summarize our findings and point out the directions of future research. 
    
\end{itemize}

\section{\label{sec:preq}Prerequisites and Search Methodology}

The idea of the hierarchical search is straightforward; the flat search algorithm is divided into two stages, \textit{stage 1} and \textit{Stage 2}, such that the number of matched-filter operations reduces successively. stage 1 search ensures matched filtering of the data sampled at the lower sampling rate ($512$ Hz) using a sparsely sampled bank called the coarse bank. Having fewer templates in a coarse bank significantly reduces the computational cost of matched filtering. Further reduction in the computation is achieved by sampling data at a lower rate than the value used in the flat search. 
The coarse bank can reduce the SNRs for a good fraction of events because of the sparsely placed templates. To compensate for the loss in SNRs, we identify triggers in each detector above {\em coarse thresholds}, set at lower values than those used in the flat search. 
We then perform a coincidence test on these identified triggers, using optimal detection statistics and obtain the zero-lag (or foreground) candidates. These foreground candidates are then followed up in stage-2 to ascertain whether they are signals or false alarms. 
%

In stage 2, we again perform matched filtering over the data segments containing followed-up foreground candidates from the stage-1 search. These data segments are sampled at a flat search sampling rate ($2048$ Hz) and filtered using a dynamic union of nbhds of mismatch extending up to $0.75$ around the followed-up stage-1 trigger templates. We refer to this union of nbhds as the \textit{stage-2 bank} from now on. The triggers generated for each detector in this stage are identified above \textit{fine thresholds}, equal to the thresholds set for SNRs in the flat search. These triggers are then subjected to a coincidence test before generating the final list of foreground candidates.

\begin{figure}[ht]
    \centering
    \includegraphics[scale=.10]{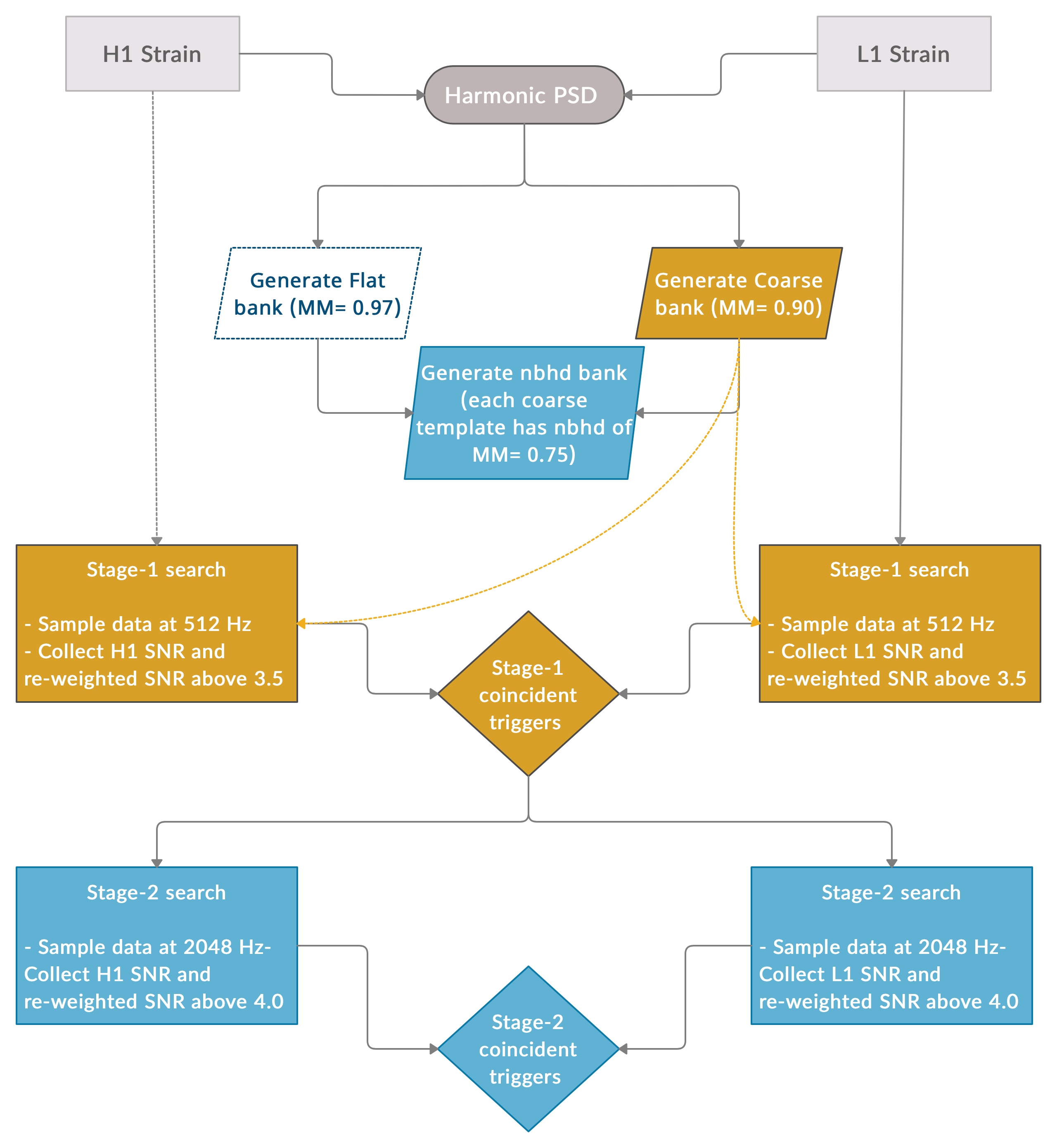}
    \caption{A flowchart depicting the working of a two-stage hierarchical search pipeline. The choice of color describes the stage: yellow for stage 1 and blue for stage 2. The first step generates the harmonic-mean power spectral density (PSD) using the strain data from the two detectors. The generated PSD is used to create flat (in dotted box) and coarse banks. Using these two banks, a nbhd bank is thus constructed. Stage-1 search begins with matched filtering of the strain data from the two detectors using a coarse bank. The generated triggers are then identified above SNRs and reweighted SNRs of 3.5. Next, a coincidence test is made to collect the foreground candidates (in diamond box), which are then followed up in stage 2 for a finer search. In stage 2, a search over the segments containing these followed-up candidates is performed using a subset of nbhd bank, stage-2 bank. The triggers generated are then identified above SNR and reweighted SNR of 4. At last, the selected triggers are then subjected to a coincidence test to obtain a final list of foreground candidates.}
    \label{fig:flowchart}
\end{figure}

Our two-stage hierarchical search pipeline is described through the flowchart in Fig.~\ref{fig:flowchart}.

\subsection{\label{subsec:templatebank}Template banks}
One of the most crucial steps in a template-based search is to adequately grid up the parameter space. A pragmatic approach suggests densely populating the search space to minimize the loss in SNR. However, such dense placement of templates makes the search computationally expensive and limits the volume and dimensionality of the parameter space that can be covered, given a fixed amount of computation power. Generally, the templates are placed such that the ``match" ($\mathcal{M}$) does not fall below a certain minimum value called the minimal match (MM). For instance, if MM is chosen as $0.97$, it means that the expected SNR for a signal does not fall more than 3$\%$ ($1 - \rm{MM} = 0.03$), corresponding to a loss of $\sim10\%$ ($\approx1 - \rm{MM}^3$) in the astrophysical events. 

The match between two normalized templates is their scalar product, maximized over the extrinsic parameters, namely, the time $t_c$ and phase  $\phi_c$ at coalescence. If $h(t_c, \phi_c, \vec{\theta})$ and $h(t^{'}_{c},\phi^{'}_{c},\vec{\theta^{'}})$ are normalized templates defined by the intrinsic parameters $\vec{\theta}$ and $\vec{\theta^{'}}$, where $t^{'}_{c} = (t_c + \Delta t_c),~\phi^{'}_{c} = (\phi_c + \Delta \phi_c),~\vec{\theta^{'}} = (\vec{\theta}+\Delta\vec{\theta})$ for $\vec{\theta} = \{m_1, m_2, s_{1z}, s_{2z}\}$, then the match is
\begin{eqnarray}
    \mathcal{M}(\vec{\theta}, \Delta \vec{\theta})=\max_{\Delta t_c, \Delta \phi_c} ~ (h(t_c, \phi_c, \vec {\theta}), h(t^{'}_{c},\phi^{'}_{c},\vec{\theta^{'}})) \,,
\end{eqnarray}
where the scalar product of arbitrary data trains $x(t)$ and $y(t)$ is defined as 
\begin{equation}\label{subeq:scalarprod}
(x, y) := 4 ~\mathcal{R}\bigg\{
    \int^{f_{high}}_{f_{low}} \frac{\tilde{x}(f) \tilde{y}^{*}(f)}{S_n{(f)}} \,df \bigg\}   \,.
\end{equation} 
Note that the match does not depend (or weakly depends) on the absolute values of the extrinsic parameters $t_c$ and $\phi_c$, and hence they have been dropped as arguments of $\M$.

In Eq.~\eqref{subeq:scalarprod}, $\mathcal{R}$ denotes the real part of a complex quantity evaluated under the sensitive frequency band, i.e., $f_{low}$ to $f_{high}$ of the detector and weighted by the detector's one-sided noise PSD $S_n{(f)}$ defined by:
\begin{equation}\label{subeq:psdeq}
     \langle \tilde{n}(f) \tilde{n}(f') \rangle= \frac{1}{2} S_n(f)\delta(f-f') \,.
\end{equation}
The angular brackets denote the ensemble average of the noise ($n(f)$) realizations. The tilde in Eqs. ~\eqref{subeq:scalarprod} and ~\eqref{subeq:psdeq} represents Fourier transform of the quantity in question, e.g., $\tilde{x}(f)$ is the Fourier transform of $x(t)$ and is given by: 
\begin{equation}\label{subeq:fft}
    \tilde{x}(f)= \int^{\infty}_{-\infty} x(t) e^{-2\pi ift}\,dt \,.
\end{equation}
Assuming a slowly varying metric $g_{mn}(\vec{\theta})$ around the targeted templates, we Taylor expand 
$\M (\vec{\theta}, \Delta \vec{\theta})$ to the lowest order of $\Delta \theta$ as:
\begin{equation}\label{subeq:1}
  \M (\vec{\theta}, \Delta\vec{\theta}) \approx 1-g_{mn}(\vec{\theta})\Delta{\theta}^m \Delta{\theta}^n \,.
\end{equation}
where the parameter-space metric is defined as:
\begin{equation}\label{subeq:2}
    g_{mn} = -\frac{1}{2} \frac{\partial^2 \M}{\partial {\Delta \theta}^m \partial {\Delta \theta}^n}({\vec{\theta}}).
\end{equation}

Therefore, by varying the source parameters $\vec{\theta}$ and calculating the metric $g_{mn}$, templates can be effectually placed in the bank. However, generally, the metric does not have a closed-form expression for aligned-spin waveforms having inspiral, merger, and ringdown phases for a wide range of source parameters, e.g., \verb+SEOBNR+~\cite{seobnr}. Moreover, in some cases, metric changes rapidly across the parameter space, making the sphere-covering problem~\cite{spherecov1} highly nontrivial. 
Therefore, techniques like stochastic placement~\cite{stochasticbank} are used to construct the bank, where a direct match is computed between templates for varying source parameters. This technique efficiently places the templates in a bank. However, if the volume of the parameter space (as defined via the metric) is large, then the template bank also becomes large and increases the computational cost for bank generation. In such a case, techniques like hybrid geometric-random placements~\cite{soumen,soumen2} efficiently generate a full nonprecessing bank. 


The density of templates in a bank relies on time-average noise PSD across all the detectors. Since the search pipeline uses a common template bank for all the detectors, a time-averaged noise PSD for each detector is estimated. These time-averaged PSDs are then combined as a harmonic mean~\cite{ianharrycwb,Keppel1,Keppel2} for the bank's construction.
%

In this work, we construct a coarse and nbhd bank for targeting GW signals from nonprecessing sources with quasicircular orbits, using Advanced LIGO-Virgo noise PSD as used in GWTC-2~\cite{gwtc2}. 
We describe the construction of banks for the parameter ranges provided in Table~\ref{table:table1} in the following sections. 



\begin{table}[ht]
\centering
\caption{Table summarizing the minimal match values and the ranges of the source parameters for the coarse, nbhd, and flat banks. The $\chi_{BH}$ and $\chi_{NS}$ are the dimensionless effective spins for a black hole and neutron star, respectively.}
\begin{tabular}{c c c c c c}
\hline
Bank & MM &$M_{tot} (M_\odot)$ &$\chi_{BH}$ &$\chi_{NS}$ & $f_{\rm min}$ (Hz)\\
\hline
Coarse & 0.90  & 2--500 & -0.998--0.998 & -0.05--0.05 &  15  \\
\makecell{Flat \&\\ nbhd} & 0.97 & 2--500 & -0.998--0.998 & -0.05--0.05 &  15 \\
\hline
\end{tabular}
\label{table:table1}
\end{table}

\subsubsection{Coarse bank}\label{subsec:coarsebank}
\begin{figure}[ht]
    \centering
    \includegraphics[scale=.26]{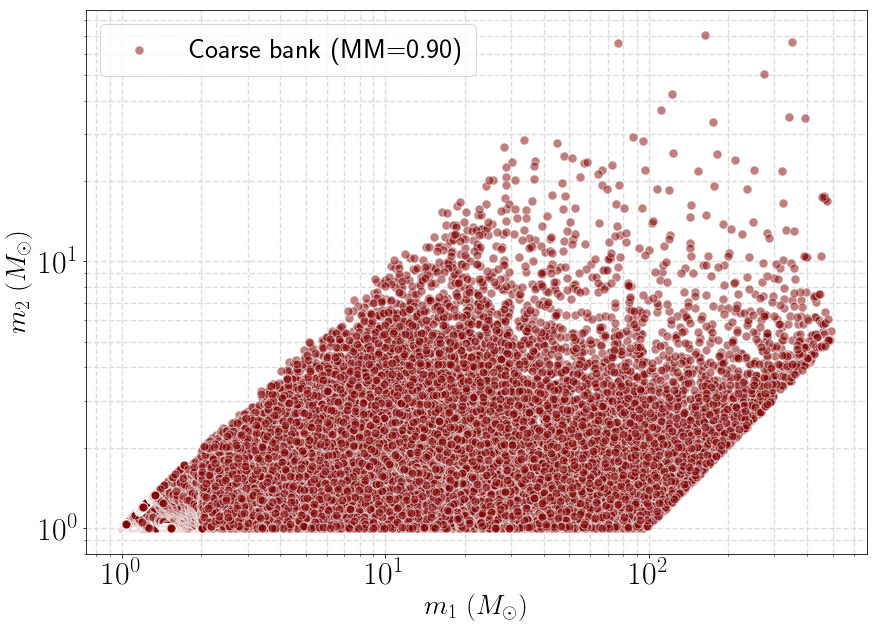}
    \caption{The distribution of coarse bank templates in $m_1 - m_2$ mass plane. Each plot point represents a template with an $MM = 0.90$ with the neighboring templates.} 
    \label{fig:coarsebankimg}
\end{figure}

We construct a coarse bank with a mismatch of $10\%$ (or ${\rm MM} = 0.90$) following~\citet{bug}, using the hybrid geometric-random method~\cite{soumen,soumen2}. The templates in the bank are generated at a minimum frequency of $15$ Hz. We discard the templates with a duration of less than $150$ ms to avoid artifacts in the matched-filtering steps. The bank is designed to search for quadrupolar, quasicircular, and nonprecessing CBC sources with the redshifted total mass ($M_{tot}$) of the binary in the range $[2~M_\odot, 500~M_\odot]$. We restrict the primary ($m_{1}$) and secondary ($m_2$) mass observed in the detector's frame in the ranges $[1~M_\odot,500~M_\odot]$ and $[1~M_\odot,120~M_\odot]$, respectively. The ranges for individual dimensionless spins of the binaries comprising a black hole ($\chi_{BH}$) and a neutron star ($\chi_{NS}$) are provided in Table~\ref{table:table1}. Thus, we construct a nonprecessing coarse bank (see Fig.~\ref{fig:coarsebankimg}) consisting of 85,080 templates. 

\begin{figure}[ht]
    \centering
    \includegraphics[scale=.27]{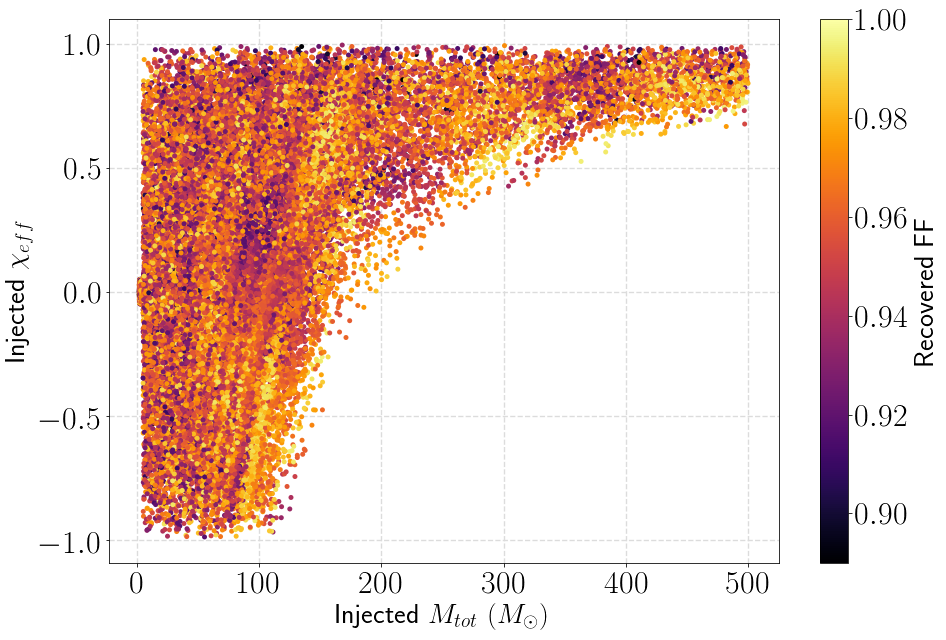}
    \caption{Recovered fitting factor as a function of effective spin ($\chi_{eff}$) and redshifted total mass ($M_{tot}$) plot for injected signals with $M_{tot}$ in the range $[2~M_\odot,500~M_\odot]$ with the signal duration cutoff of 150 ms.}
    \label{fig:ff_broad_20k_bbh}
\end{figure}

To check whether the bank does not possess holes in the parameter space, we test the bank's performance in terms of fitting factor (FF)~\cite{fittingfactor}. In this test, we estimate FF for $\sim80,000$ quasicircular, quadrupolar, spin-aligned, and nonprecessing CBC signals that span the bank's search parameter space. We use \verb+TaylorF2RedSpin+~\cite{chinm} with $M_{tot}$ in the range $[2~M_\odot,5~M_\odot]$ and $|\chi_{eff}|\leq0.05$, and \verb+SEOBNRv4_ROM+~\cite{seobnr} in $[5~M_\odot,500~M_\odot]$ with $|\chi_{eff}|\leq0.998$. We recover FF greater than 0.90, as can be seen in Fig.~\ref{fig:ff_broad_20k_bbh}. This result signifies that our bank is effectual and suffices the design criteria as per Table~\ref{table:table1}.

\subsubsection{Neighborhood bank}\label{subsec:nbhd}
For a template corresponding to stage-1 coincident trigger, the template nbhd is the region in parameter space where mismatch with neighboring templates can be up to $25\%$, as described in Sec. IIIB2 of~\citet{bug}. To sample these nbhds, we use a pregenerated flat bank with $MM=0.97$ with the search parameter space provided in Table~\ref{table:table1}. We include flat bank's templates having ${\rm MM} \ge \MMn \equiv 0.75$ with the trigger template. We calculate nbhds for all the coarse templates. This precomputed bank with assigned nbhds is referred to as a nbhd bank, and a dynamic subset of it is termed as a stage-2 bank. The stage-2 bank is dynamic because the number of templates residing in this bank changes depending on the noise realization of each segment. 


To identify the nbhds of coarse templates, we adopt the following strategy. For coarse templates with $M_{tot}\geq 12$ $M_\odot$, we perform an exact match calculation with all the flat bank templates. For templates with $M_{tot} < 12$ $M_\odot$~\cite{chinm}, we first shortlist a set of templates that may be able to satisfy the nbhd criteria. For that, we define a minimal match ellipsoid with ${\rm MM} = \MMn$ in the following way:
Consider a coarse template of $M_{tot} < 12$ $M_\odot$ for which nbhd has to be calculated. We first construct a minimal-match ellipsoid centered at this template in a coordinate system where the metric varies slowly over the parameter space, i.e., the metric is almost constant, and the signal manifold is almost flat. Therefore we choose chirp-time coordinates \{$\tau_0$, $\tau_3$, $\tau_{3s}$\}, collectively labeled as $\tau^\a$. These coordinates are given by scaling \{$\theta_0$, $\theta_3$, $\theta_{3s}$\}, described in Ref.~\cite{soumen}, with $(2\pi f_{o})^{-1}$ at $f_o = 20$ Hz. In these coordinates, we estimate the metric components using \verb+TaylorF2RedSpin+ waveform model. Once the metric is known, we, following~\cite{soumen}, diagonalize it by an orthogonal transformation $\mathcal{O}$, and obtain the eigenvalues $\gamma_\a$ with new coordinates $\xi^\a = \mathcal{O}^\a_\b \tau^\b$. The metric in these coordinates is in a diagonal form and is given by
\begin{equation}
ds^2 = \sum_{\a = 1}^3 \gamma_\a (d \xi^\a)^2    \,.
\end{equation}
This is just a principal axis transformation to an orthogonal basis. Along the eigendirections, the lengths of the semiaxes [$r_\a (\MM)$] of the ellipsoid for a given value of ${\rm MM} $ are given by 
\begin{equation}\label{subeq:raxes} 
    r_\a (\MM) = \sqrt{\frac{1- \MM}{\gamma_\a}} \,.
\end{equation}
As $\MM$ reduces from its maximum value of unity, the ellipsoid increases in size. In the $\xi^\a$ coordinates, let the coarse and flat templates be labeled by $\xi_0^\a$ and $\xi^\a$, respectively. Let $\Delta \xi^\a = \xi^\a - \xi^a_0$, and define the distance $d(\xi^\a, \xi^\a_0)$ by the equation
\begin{equation}\label{subeq:dist}
    d^2 (\xi^\a, \xi^\a_0) = \sum_{\a = 1}^3 \gamma_\a (\Delta \xi^\a )^2.
\end{equation}
Then, the relation $d(\xi^\a, \xi^\a_0) \leq \sqrt{1 - \MMn}$ defines the ellipsoid in $\xi^\a$ coordinates. We use this ellipsoid to guide our selection of flat templates. Note that the metric approximation is extrapolated beyond its validity regime, so the ellipsoid is only a crude estimate of the nbhd. In any case, since we have made a conservative choice of $\MM = \MMn$, we do not expect to miss out on any signals. We choose templates accordingly in this region and compute the match between a flat template and a given coarse template. If the match is above the stipulated $\MMn$, we retain the template in the nbhd. \textit{Thus, the final list of templates in the nbhd is obtained by the actual computation of the match between coarse and fine templates inside the ellipsoid.}  

\begin{figure}[ht]
    \centering
    \includegraphics[scale=.25]{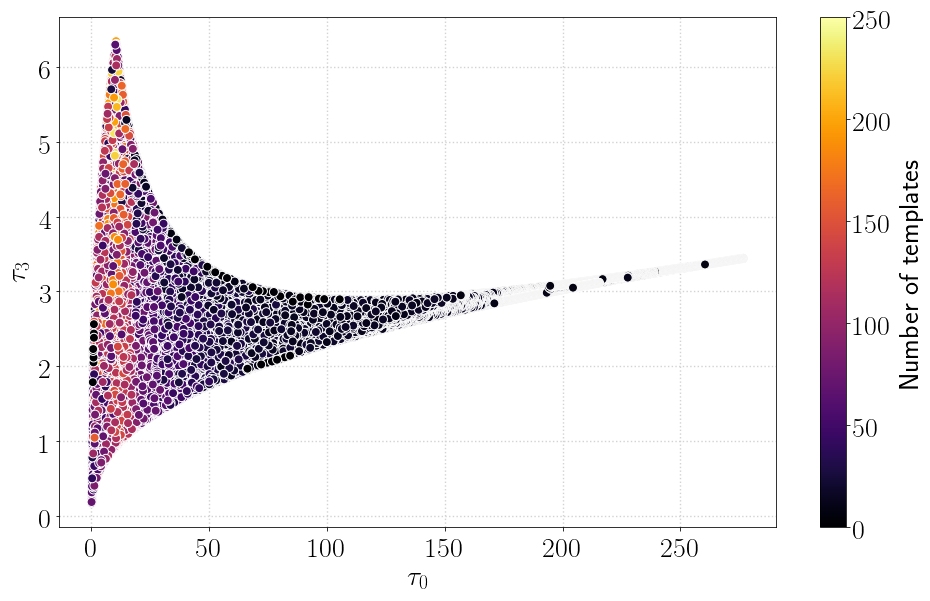}
    \includegraphics[scale=.23]{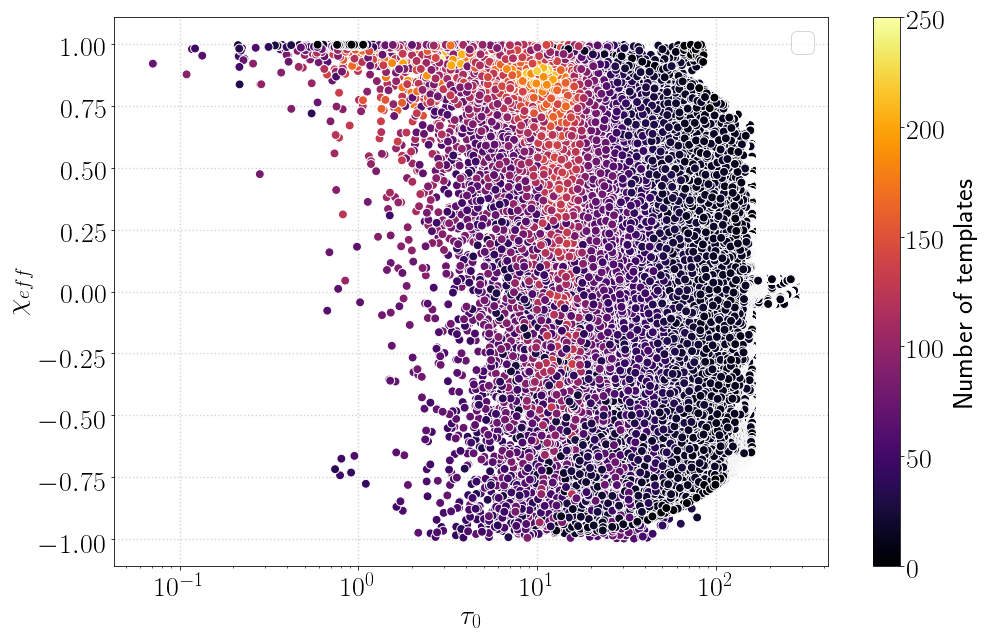}
    \caption{The distribution of nbhd bank templates in $\tau_{3} - \tau_{0}$ (top) and $\chi_{eff}-\tau_0$ plane (bottom). The color scale represents the number of templates in the nbhd of each coarse template. Typically, there are $\sim40 - 150$ flat templates in the nbhd of a coarse template.}
    \label{fig:stage2bank}
\end{figure}

\begin{figure}[ht]
    \centering
    \includegraphics[scale=.24]{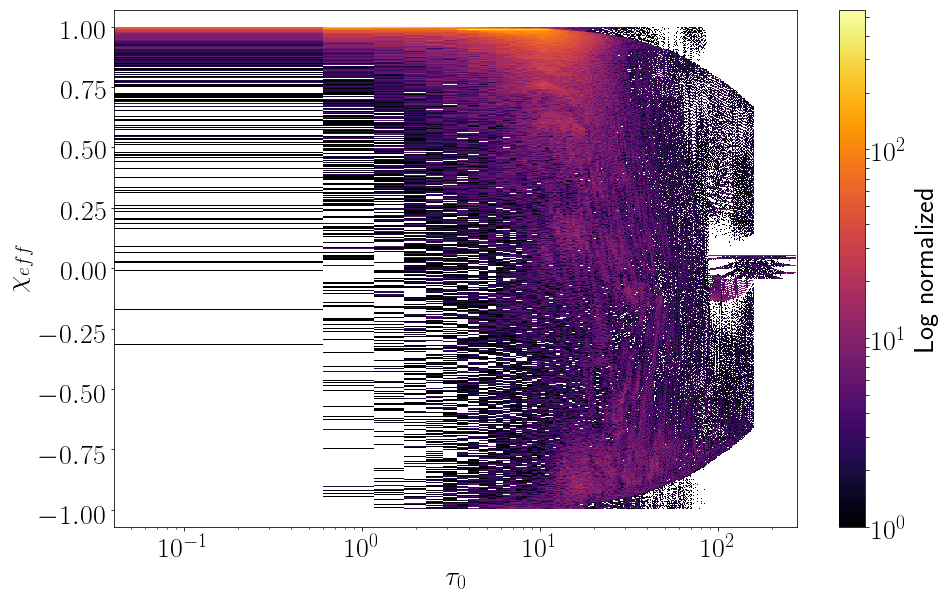}
    \caption{Plot showing the density of flat bank's templates in $\chi_{eff}-\tau_0$ plane. The color scale represents the log-normalized number density.}
    \label{fig:flatbank}
\end{figure}

In general, we find that a single nbhd around a coarse template (not very close to the boundary of the parameter space) contains $\sim40 - 150$ templates. Since the match falls gradually with an increasing mismatch in the $\tau_3$ mass parameter (as compared to $\tau_0$), the nbhd tends to extend considerably along with this coordinate (see Fig.~3 in \citet{anandsen}). Therefore, a large portion of the nbhd can extend outside the physical parameter space considered for the search, especially for higher $\tau_{0}$. This causes a significant variation in the number of templates in the nbhd, as reflected in the top panel of Fig.~\ref{fig:stage2bank}. It is also interesting to note that the variation in the number of templates in the nbhd (bottom panel of Fig.~\ref{fig:stage2bank}) resembles the actual template density of the flat bank plotted in $\tau_0$ and $\chi_{eff}$ coordinates (Fig~\ref{fig:flatbank}). The figure indicates that there is a higher template density around high $\chi_{eff}$ and low $\tau_0$. 

%

\subsection{\label{sec:matchedfilter}Matched filter}
The model-dependent search for GWs from CBCs using templates in the LIGO-Virgo data exploits the matched filtering~\cite{svd} technique rigorously. This technique correlates discretely sampled time-series data $s(t)$ with the normalized templates $h(t_c,\phi_c,\vec{\theta})$ for the source parameters ($\vec{\theta}$) within the detectors' sensitive band. The correlation generates matched-filter SNR time series $\rho(t_c)$ maximized over the coalescence phase $\phi_c$, and it is defined as:
\begin{equation}\label{subeq:snr}
     \rho \left( t_c; {\vec{\theta}} \right) \equiv 
    \left| \left( s, (1+i)h({ t_c,\phi_c = 0,\vec{\theta}})\right) \right| \,.
\end{equation}
Generally, the data obtained from the detectors are non-stationary and non-Gaussian~\cite{detchar1,detchar2,detchar3}. Preprocessing steps involving data-quality checks and application of vetoes flag most of the artifacts present in the data~\cite{dqpaper}. Nevertheless, the short-duration glitches or long-duration correlations, as described in~\citet{venumadhav1}, still remain in it. Matched filtering over these noise transients often leads to high SNRs. These short-duration noise transients are removed from the standard search pipelines by nullifying noise contributions in the time-series data via gating~\cite{usman}. We, therefore, apply a similar gating priory to the matched filtering for each detector to remove the nonstationary transients from the strain data in our analysis. 

Matched filtering the data produces several triggers with varying SNRs for each template in the bank. These triggers are first clustered within a time window of 0.5 s to retain only the ones with high SNRs~\cite{findchirp}. In the second step, the SNRs of triggers due to loud noise artifacts are suppressed using signal consistency tests like the standard chi square ($\chi_{r}^2$)~\cite{usman,allen}, and sine-Gaussian chi square ($\chi_{sg}^2$)~\cite{nitzsgveto}. 

Like in flat search, the trigger SNRs ($\rho$) generated in both the stages of the hierarchical search are down-weighted with their reduced chi-square values using $\chi_{r}^2$~\cite{usman,allen} veto defined as
\begin{align}
\label{subeq:newsnr}
\tilde{\rho} = \begin{cases}
\frac{\rho}{[(1+(\chi_r^2)
^{3})/2]^{1/6}} & \text{if $\chi_r^2> 1$}, \\
\rho & \text{otherwise}.
\end{cases}
\end{align}
Usually, $\chi^2_r$ veto is ineffective in the region where signals are too short. In such cases, the short-duration templates ring with ``blip" glitches present in the data. Therefore, we further down-weight $\tilde{\rho}$ for the templates with $M_{tot}> 30~M_\odot$ using $\chi_{sg}^2$~\cite{nitzsgveto} veto defined as
\begin{align}
\label{subeq:newsnrsgveto}
\hat{\rho} = \begin{cases}
\tilde{\rho}~{(\chi^2_{r,sg})}^{-1/2} & \text{if $\chi^2_{r,sg} > 6$}, \\
\tilde{\rho} & \text{otherwise}.
\end{cases}
\end{align}
In each stage of the hierarchical search, the triggers that surpass the two tests above specific thresholds on $\rho$ and $\tilde{\rho}$ (see Sec. \ref{sec:pipeline}) are subjected to a coincidence test to recover the real GW events. The coincident events are obtained based on the optimal detection statistics as defined in~\cite{nitzphasetd,gareth}, which we elaborate on in the following section. 
\subsection{\label{sec:ranking} Ranking statistics}
A pair of triggers from the two detectors is coincident if it simultaneously occurs within the light travel time between them and is recovered with identical template parameters. The coincidence is evaluated based on optimal detection or ranking statistics ($\Lambda_{opt}$), defined as the ratio of the likelihood for data containing signal to the likelihood for data having noise~\cite{detectionstatistic}. These likelihoods are the functions of the template parameters ($\vec{\theta}$) and $\hat{\rho}$, $\chi_{r}^2$. 

In the recent works~\cite{gareth,gwtc1,ogc1,ogc2}, the optimal detection statistics were approximated by taking the ratio of coincident event rate densities due to signal ($p(\vec{\kappa}|S)$) and noise $(p(\vec{\kappa}|N))$. Therefore, for an unknown coincident with template parameters $\vec{\kappa}=\{\hat{\rho}_{H},\hat{\rho}_{L}, \chi^2_{H},\chi^2_{L},\delta t_c, \delta\phi_c, \vec{\theta}\}$ where $\delta t_c,~\delta\phi_c$, is the time and phase difference in between two detectors, $\Lambda_{opt}$ is given as
\begin{equation}\label{eq:rankingstat}
    \Lambda_{\rm opt} =\frac{p(\vec{\kappa}|S)}{p(\vec{\kappa}|N)} \, \equiv \frac{p(\vec{\kappa}|S)}{r^{HL}_{\vec{\theta}} \: p(\vec{\theta}, \delta t_c, \delta \phi_c|N)} \,.
\end{equation}
For the statistics, $p(\delta t_c, \delta \phi_c|N)$ is expected to be uniform over $(\vec{\theta}, \delta t_c, \delta \phi_c)$~\cite{detectionstatistic}; thereby it is marginalized and treated as a constant. If the noise is uncorrelated between detectors, $p(\vec{\kappa}|N)$ ($\approx r^{HL}_{\vec{\theta}}$) can be safely written as a product of single-detector noise rate densities~\cite{gareth} ($r_{\vec{\theta},H},~r_{\vec{\theta},L}$) given by
\begin{equation}
    r^{HL}_{\vec{\theta}}= 2~\tau_{HL} (\: r_{\vec{\theta},H} (\hat{\rho}_H) \: r_{\vec{\theta},L} (\hat{\rho}_L)) \,,
\end{equation}
where, $\tau_{HL}$ is the allowed time window for a coincidence of trigger in twin LIGO detectors at Hanford (H) and Livingston (L).

Thus, by estimating $r^{HL}_{\vec{\theta}}$ and $p(\vec{\kappa}|S)$ through accurate modeling~\cite{nitzsgveto,gareth}, one can obtain $\Lambda_{opt}$ for the coincident triggers.

In each stage of the hierarchical search, we model $r^{HL}_{\vec{\theta}}$ and $p(\vec{\kappa}|S)$ separately to obtain the ranking statistics of coincident and time-shifted events. In the first stage (and flat search), we adopt a similar methodology of modeling coincident signal and noise rate densities for a two-detector configuration as described in~\citet{gareth}. However, we model coincident noise rate density slightly differently in the second stage.
In the following sections, we first review the existing modeling procedure for signal and noise rate densities used by the flat and stage-1 search and then elaborate on modeling noise rate densities for stage-2 search.

\subsubsection{\textbf{Signal model:} For flat, stage 1, and stage 2}\label{subsec:signalmodel}
To model $p(\vec{\kappa}|S)$, one requires the probable astrophysical distribution of the binary sources that Advanced LIGO detectors can detect. In reality, the exact distribution is unknown to the observers. Nevertheless, the source population can be approximated as uniform in volume and isotropic in the sky location and orientation of the binary. Assuming these distributions for sources, we can estimate how their detection parameters like signal amplitudes, time, and phase differences vary with respect to the pair of the LIGO detectors. 

As described in~\cite{nitzphasetd,gareth}, $p(\vec{\kappa}|S)$ is precomputed by performing Monte Carlo simulations assuming fixed detector sensitivity. Then the corresponding $p(\vec{\kappa}|S)$ is used to rank each coincident trigger with parameter closed to $\vec{\kappa}$~\cite{nitzphasetd}.

We use the above recipe to generate $p(\vec{\kappa}|S)$ in the flat and both stages of the hierarchical search. 

\subsubsection{\textbf{Noise model:} For flat and stage 1\label{subsec:stage1ranking}}
The coincident noise event rate density, $r^{HL}_{\vec{\theta}}$, for the flat and stage-1 search is obtained by first estimating the single-detector noise rate densities ($r_{\vec{\theta},d}|_{d=\{H,L\}}$). Like in~\citet{gareth}, this quantity in the flat and stage-1 search is calculated by modeling the tail of the trigger distribution for each detector ($d$) and template with a falling exponential function as
\begin{equation}\label{subeq:singledetrates}
    r_{\vec{\theta},d}(\hat{\rho}_{d},N)= \mu(\vec{\theta})~p(\hat{\rho}_{d}|\vec{\theta},N),
\end{equation}
given, 
\begin{align}
\label{subeq:exponential}
p(\hat{\rho}_{d}|\vec{\theta},N) = \begin{cases}
\alpha(\vec{\theta}) \exp[-\alpha(\vec{\theta})(\hat{\rho}_{d}-\hat{\rho}_{th,d})] & \text{if $\hat{\rho}_{d} > \hat{\rho}_{th,d}$}, \\
0 & \text{otherwise},
\end{cases}
\end{align}
where $\mu(\vec{\theta})$ and $\alpha(\vec{\theta})$ denote trigger count above the threshold ($\hat{\rho}_{th,d}$) and exponential decay rate, respectively. 

The  fit parameter $\alpha(\vec{\theta})$ is obtained by maximum logarithmic likelihood fitting method. For discrete samples of $\hat{\rho}_{d}$ of $k$th trigger, we maximize
\begin{equation}\label{subeq:loglikeli}
    \ln p(\hat{\rho}_d | \alpha, n) = n \ln \alpha -\alpha \sum^{n}_{k} (\hat{\rho}_{k,d} - \hat{\rho}_{th,d}),
\end{equation}
at a fixed $\hat{\rho}_{th,d}$ ($\equiv 6$) to obtain $\alpha_{max}=(\bar{\hat{\rho}}_{d}-\hat{\rho}_{th,d})^{-1}$. Here, $\bar{\hat{\rho}}_{d}$ is the mean of $\hat{\rho}_{d}$ and the variance ($\sigma_{d}$) in fit parameter is given by $1/\sqrt{n}$, where $n$ denotes the number triggers generated for a particular template.

In the flat and stage-1 search, we calculate $\alpha_{max}$ and $~n$ for each flat and coarse template, respectively. Generally, not all the templates have sufficient triggers above $6$ to fit the trigger distribution's exponential tail. In such cases, the low number of triggers gives a high variance to the fit parameter values. To avoid such problems, we take the moving average of the fit parameters and smooth $\mu(\vec{\theta})$ by taking mean over nearby templates with similar values of effective spin, template duration, and symmetric mass ratio as performed in~\citet{gareth}.  

\subsubsection{\textbf{Noise model:} For Stage 2}\label{subsec:noisemodelstage2}
In principle, the procedure for obtaining single-detector noise rate densities described previously can be applied in stage 2. However, it cannot be implemented, as this stage possesses insufficient triggers above $\hat{\rho}_{th,d}$ to obtain meaningful fit parameters. The reason is, we follow only foreground candidates from stage 1 that have $\Lambda_{opt}\geqslant 7$. Matched filtering over these followed-up triggers utilizes fewer nbhds and corresponding templates to generate fewer triggers. Having an inadequate and biased set of triggers for a template can give a significant variance in the values fit parameters, leading to overestimating single-detector noise rates if we only use stage-2 triggers. We, therefore, \textit{do not} explicitly calculate the fit parameters in stage 2. Instead, we reuse the fit-values of the ``closest" coarse template to the stage-2 trigger template. The ``closeness" relies on the highest match value between the coarse and stage-2 bank templates. 

\begin{figure}[ht!]
    \centering
    \includegraphics[scale=.23]{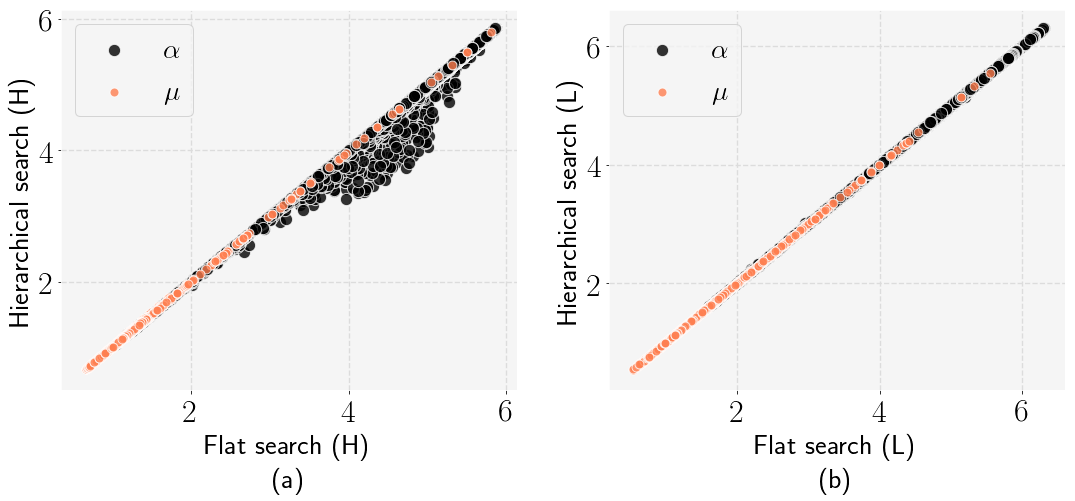}
    \caption{Comparison plot for the fit coefficients, $\alpha$ and $\mu$, obtained from the flat and hierarchical search for (a) Hanford (H) and (b) Livingston (L) detectors.}
    \label{fig:fitcomp}
\end{figure}

To verify the applicability of the above procedure, we perform a flat and hierarchical search on 14 days and obtain fit parameters. Figure~\ref{fig:fitcomp} compares the fit parameters obtained in both the searches. The scatter points in the diagonal signify that the values are comparable for the two searches in both the detectors. A few templates in the Hanford detector show low $\alpha$ indicating small fluctuations in their values due to noise. These small fluctuations can appear at different periods of observational time. However, these variations in $\alpha$ negligibly affect the modeling of single-detector noise rate density, as can be seen later in Sec.~\ref{sec:comparison}.

\subsection{Assigning significance\label{sec:significance}}

The significance of any event is evaluated based on their FAR estimate above a fixed statistic $\Lambda^{*}$ as:
\begin{equation}\label{eq:FAR}
    \text{FAR}(\Lambda^{*}) = \int d^n \vec{\kappa}  ~r^{HL}_\kappa \Theta(\Lambda_{opt}(\vec{\kappa})-\Lambda^*)\,,
\end{equation}
where $r^{HL}_{\vec{\kappa}} \equiv r^{HL}_{\vec{\theta}}$ by construction. 
False Alarm Rate (FAR) signifies the rate of occurrence of a nonastrophysical coincident candidate with a similar or higher $\Lambda_{opt}$ [see Eq.~\eqref{eq:rankingstat}] in the observing period. FAR is estimated in the flat and stage-1 search with respect to a noise background constructed by time-sliding data by a minimum of 100 ms across the detectors. Such a procedure omits all the possibilities to have a coincidence due to a real GW signal. At each time shift, $\Lambda_{opt}$ is recomputed to rank the candidates above a certain threshold ($\Lambda^{*}$). 
Performing several time shifts generates many plausible candidates that could be cumbersome to store. In order to mitigate the storage problem, the background computation is optimized in the standard PyCBC search. At first, a clustering over time is performed such that candidates with the highest statistic value, falling within 10 s, are kept. In the next step, candidates are selectively chosen with all or few time slides falling in the ranking statistic value's bin. For instance, candidates with all possible time slides with ranking statistics greater than 9 are chosen, but only some are selected with time slides of 30 s for which statistic value lies between 8 and 8.5.

In principle, a similar strategy can be implemented to assign FARs to the detected candidates in stage 2 of the hierarchical search. However, the background constructed by time-sliding stage-2 triggers using a union of stage-2 banks can bias the detected candidates' FAR estimates, as shown in~\citet{bug}. Therefore, we avoid such biases by constructing an approximate background that would mimic a background constructed in the flat search. As proposed in~\citet{bug}, we construct a scaled stage-1 background for assigning significance to the final list of coincident triggers. First, we construct a stage-1 background by time-sliding stage-1 triggers by 100 ms across the detectors as done in the flat search. We then scale this background by a factor equal to the ratio of the number of templates times the sampling frequency used in a flat search to stage 1. This number turns out to be close to the computational gain and is approximately $20$. 
%

    

\begin{figure}[ht]
    \centering
    \includegraphics[scale=.22]{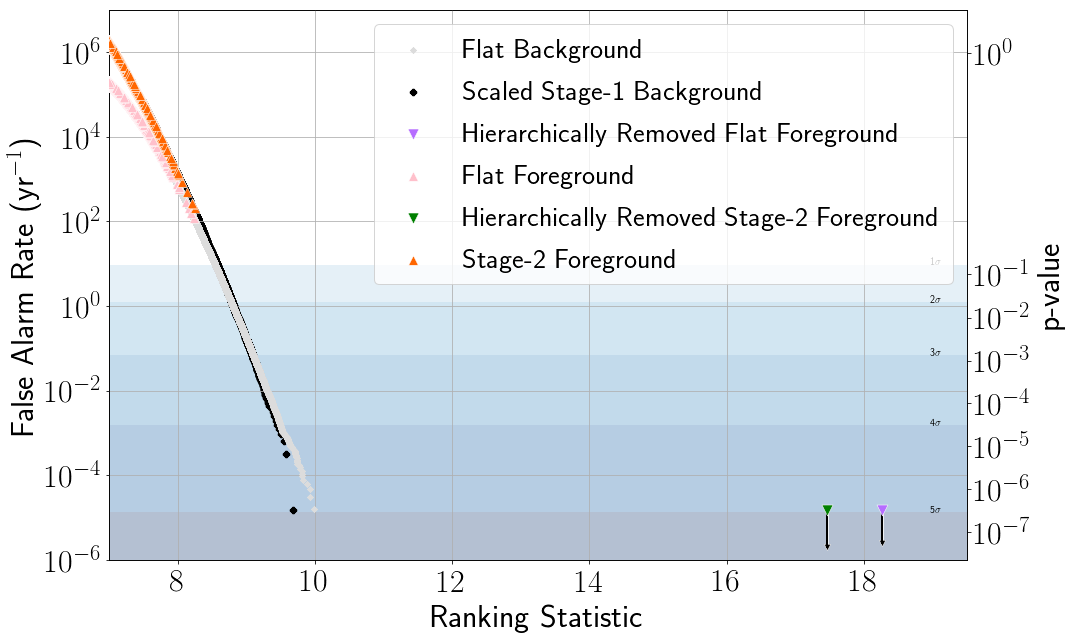}
    \caption{Plot showing FAR vs. ranking statistic curves for the foreground candidates (foreground) and the time-shifted candidates (background) from a flat and hierarchical search. The foreground (triangle) overlays the background (circle) in each search. The loudest event, GW150814, is hierarchically removed from the background in both searches. Note that the scaled stage-1 background (black) roughly matches the flat background (gray) above ranking statistic value 8. }
    \label{fig:bgcomp}
\end{figure}

\begin{figure}[ht!]
    \centering
    \includegraphics[scale=.22]{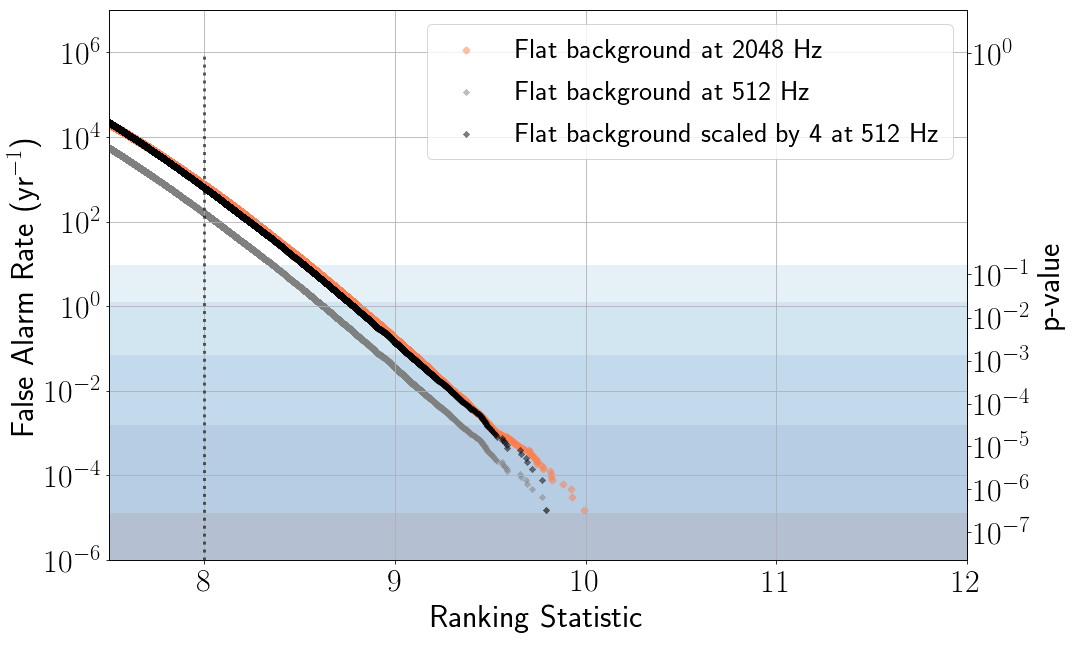}
    \includegraphics[scale=.22]{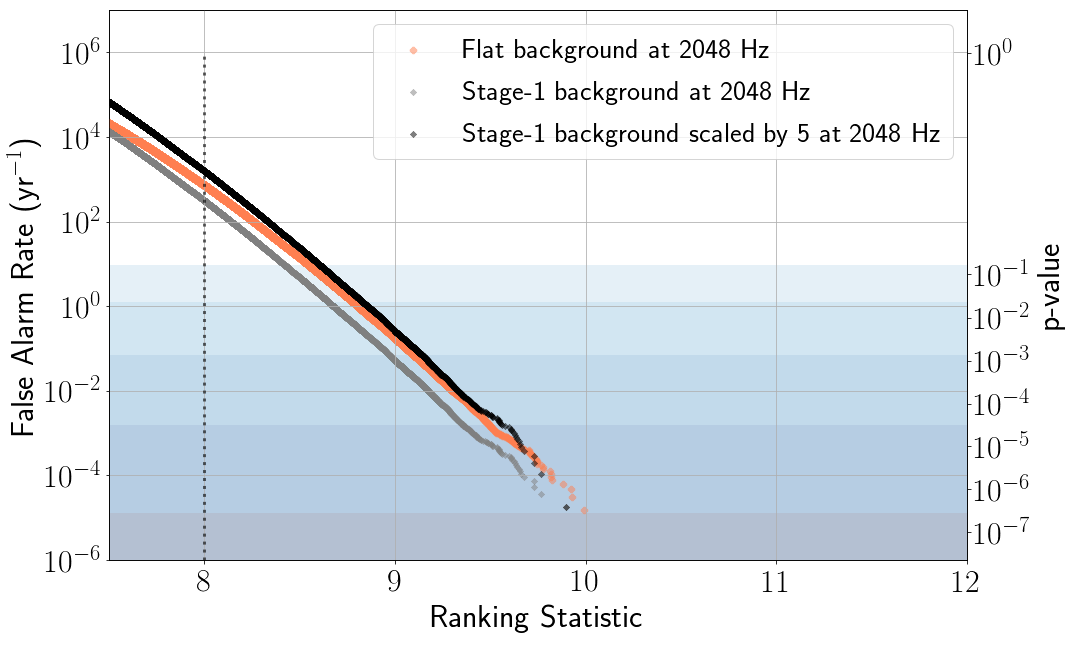}
    \caption{Plot depicting FAR vs. ranking statistic curves for the backgrounds obtained for the flat and stage-1 search at different sampling rates. The top panel shows the backgrounds obtained for the flat search performed at sampling rate- 512 Hz (gray) and 2048 Hz (orange). The bottom panel shows the background obtained at a sampling rate of 2048 Hz for the flat (orange) and stage-1 search (gray). On scaling the flat background obtained at a sampling rate of 512 Hz by a factor of 4 (top), and stage-1 background obtained at 2048 Hz by 5 (bottom), the resultant backgrounds (black) approximately match the standard flat background (orange) obtained at 2048-Hz sampling rate.}
    \label{fig:bgscaling}
\end{figure}

To justify our argument on mimicking a flat background, we compare the foreground and background obtained from the flat and hierarchical search performed over 14 days of data around the first BBH, GW150914~\cite{firstevent}, event. We find that the foregrounds due to noise candidates match their respective backgrounds for both the searches, as shown in Fig.~\ref{fig:bgcomp}. We observe that the noise background is higher in the lower ranking statistics region than that of flat. This observation is expected as the scaling factor linearly increases the number density of noise triggers in a particular ranking statistics bin. We also notice that the scaled stage-1 background roughly matches the flat background above ranking statistic value 8. Therefore, the reliability of the FARs will be limited to the ranking statistic value $\gtrsim 8$. Another way to justify the reliability of the background is by looking at the effects of variations in sampling rates and the number of templates for each pair of the search. Figure~\ref{fig:bgscaling} compares the backgrounds obtained from the flat search at 512 Hz and the stage-1 search at 2048-Hz sampling rates. We show that if these backgrounds are scaled with a factor of 4 and 5, respectively, both nearly match the standard flat search background at 2048-Hz sampling. Thus, the factor of 5 reductions in the number of templates and 4 reductions in the sampling rate, whose product gives us the final scaling factor of 20, are valid scaling factors on their own. While the scaling argument still lacks concrete proof, we think it makes our argument much more robust, at least for the standard search with a bank of mismatch 0.97.

While the scaled background may not precisely match the background of the flat search, it is still a monotonic function of the detection statistic and reasonably close to the flat search background. Hence, the FAR estimate based on the scaled background can be used for detection, as long as a reasonable FAR threshold to claim a detection is decided by comparing it with the corresponding flat search.

\section{\label{sec:pipeline}Search for CBC in O1 and O2 data}
We search for CBCs using the two-stage hierarchical search~\cite{kanchan} over the data from the first (O1) and second (O2) observing runs of twin LIGO detectors. We use $21.39$ days of coincident data from O1 and $31.4$ days from O2. 

The periods of poor-quality data are marked and removed from the analysis using data-quality flags, Category 1 (CAT-1) and Category 2 (CAT-2)~\cite{dqpaper}. CAT-1 vetoes remove the times during which at least one of the key components of a detector was not operational in the nominal configuration due to critical issues. The duration over which excessive noise is observed due to instrumental artifacts is marked and removed by CAT-2 flags. 

As described in Sec.~\ref{sec:matchedfilter}, the data undergo preprocessing before entering the matched filtering step. In both stages, we use 512 s of overlapping data segments for matched-filter computation. 
We pad data segments with zeros 144 s at the beginning and 16 s at the end to avoid the artifacts generated from the discrete Fourier transform. Once the data segment is prepared, we perform a hierarchical search in two stages. 

We begin the search by matched-filtering data segments sampled at $512$ Hz with a coarse bank (see Sec.~\ref{subsec:coarsebank}) and obtain a list of stage-1 triggers above coarse thresholds on $\rho$ and $\tilde{\rho}$. Triggers with $\rho>3.5$ that pass the $\chi_{r}^2$ test with $\tilde{\rho}>3.5$, get further reweighted by $\chi_{sg}^2$ veto. The choice of coarse thresholds for stage-1 search may seem arbitrary. However, we tested out different values for $\rho$ and $\tilde{\rho}$ and found that setting both values at $3.5$ gives the optimal computational cost of handling bulk triggers.

The surviving single-detector triggers then undergo a coincidence test (see Sec.~\ref{sec:ranking}) to obtain foreground candidates. These foreground candidates are then followed up in stage 2. 
%

Stage 2 of the hierarchical search begins with matched-filtering data segments sampled at $2048$ Hz that contain foreground candidates with $\Lambda_{opt}\geqslant7$~\cite{bug} from stage 1. Each segment is filtered using a unique stage-2 bank (see Sec.~\ref{subsec:nbhd}) constructed from the dynamic union of the nbhds around each followed-up trigger template. 
The matched-filter SNRs generated in this stage are then reweighted with fine thresholds on $\rho$ and $\tilde{\rho}$ of 4. As described in Sec.~\ref{sec:ranking}, the resultant triggers are then subjected to a coincidence test to obtain the second stage's foreground candidates.

The final step in the search involves assigning significance to the potential foreground candidates ($\Lambda_{opt}>8$) obtained in stage 2. We assign FARs to these candidates using a scaled stage-1 background, as described in Sec.~\ref{sec:significance}. Based on this background, we present the results from the analysis in the next section. 

\begin{table*}[ht!]
       \caption{List of GW events detected via hierarchical search. The events are arranged in the ascending order of their event time. We report and compare detected events' FARs, network SNRs ($\hat{\rho}_{T}$), and redshifted chirp masses ($\mathcal{M}_{chirp}$) in the stage-1 and stage-2 search with the events reported by the flat search in~\citet{gwtc1}. We see an improvement in the FAR and network SNR values for the events, with network SNR varying between 9 and 12 from stage 1 to stage 2.}  \renewcommand{\arraystretch}{1.1}
        \begin{tabular}{c c@{\hskip 0.1in}  c@{\hskip 0.3in} 
        c@{\hskip 0.1in} c@{\hskip 0.1in}  c@{\hskip 0.3in} c@{\hskip 0.1in} c@{\hskip 0.1in} c@{\hskip 0.3in} c@{\hskip 0.1in} c@{\hskip 0.1in} c@{\hskip 0.1in}}
        
\hline\hline
 \multirow{1}{*}{Sl. no.}   & \multirow{1}{*}{Event} & \multirow{1}{*}{UTC} &  \multicolumn{3}{c}{Flat} &  \multicolumn{6}{c}{Hierarchical} \\
\cline{7-12}
&&&&&&\multicolumn{3}{c}{Stage-1} & \multicolumn{3}{c}{Stage-2} \\ 
\cline{7-12}
&  & & \makecell{FAR \\ ($\mathrm{yr^{-1}}$)} & $\mathrm{\hat{\rho}_{T}}$ & \makecell{$\mathcal{M}_{chirp}$ \\($\mathrm{M_\odot)}$} & \makecell{FAR \\($\mathrm{yr^{-1}}$)} &
$\mathrm{\hat{\rho}_{T}}$ & \makecell{${\mathcal{M}_{chirp}}$ \\($\mathrm{M_\odot)}$} & \makecell{FAR \\($\mathrm{yr^{-1}}$)} & $\mathrm{\hat{\rho}_{T}}$  & \makecell{$\mathcal{M}_{chirp}$ \\($\mathrm{M_\odot)}$} \\
\hline
& & & & & & & & & &\\
1  & GW${150914}$ & 09$:$50$:$45.4  & $\leq$ 1.53$\times10^{-5}$  & 23.6 & 32.75 & 1.52$\times 10^{-5}$ & 23.3& 29.71&  1.52$\times 10^{-5}$ & 24.0 & 31.96\\
2  & GW${151012}$ & 09$:$54$:$43.4 & 0.17 & 9.5 & 18.47 &  0.42 &8.9& 18.68& 0.05 &  9.8 & 18.31\\
3  & GW${151226}$ & 03$:$38$:$53.6 & $\leq$ 1.70$\times10^{-5}$ & 13.1 & 9.70 & 1.69$\times 10^{-5}$ &11.9 & 9.89& 1.69$\times 10^{-5}$  & 13.1 & 9.72\\ 
4  & GW${170104}$ & 10$:$11$:$58.6 & $\leq$ 1.39$\times10^{-5}$ &
13.0  & 20.19 &1.37$\times10^{-5}$  &12.2 &18.37 &  1.37$\times10^{-5}$ & 12.9 & 29.17\\
5  & GW${170608}$ & 02$:$01$:$16.5 & $\leq$ 3.09$\times10^{-4}$ &
15.4 & 8.61	& 3.08$\times10^{-4}$ & 8.9& 8.65& 3.08$\times10^{-4}$ &  14.8 & 9.03\\
6  & GW${170729}$ & 18$:$56$:$29.3  & 1.36 & 9.8 &40.27 &1.68
& 9.3& 54.41& 0.05  & 10.6 & 47.51\\
7  & GW${170809}$ & 08$:$28$:$21.8  & 1.45$\times10^{-4}$ & 12.2 & 23.53& 0.56& 11.3 &29.71 & 
1.70$\times10^{-3}$ & 12.1 & 23.65 \\
8 & GW${170814}$ & 10$:$30$:$43.5 & $\leq$ 1.25$\times10^{-5}$ & 16.3 & 25.20 & 1.25$\times 10^{-5}$ & 16.0& 25.09& 1.25$\times 10^{-5}$  & 17.2 & 26.58\\
9  & GW${170817}$ & 12$:$41$:$04.4 & $\leq$ 1.25$\times10^{-5}$ & 30.9 & 1.20& 2.51$\times10^{-5}$ & 28.7& 1.20& 1.25$\times10^{-5}$  & 31.5 & 1.20\\
10 & GW${170823}$ & 13$:$13$:$58.5  & $\leq$ 3.29$\times10^{-5}$ & 11.1 & 23.61
&  3.30$\times10^{-5}$  &11.3 & 32.32&  3.30$\times10^{-5}$  & 11.1 & 46.85 \\
\hline\hline
\end{tabular}
\label{table:table2}
\end{table*}

We report the recovery of all ten confirmed GW events with FAR below 1 per year in stage 2 of hierarchical search. These events were previously detected by the flat analysis in GWTC-1. Although the detection statistics used in both the stages of hierarchical search are more recent than those used in the flat analysis of GWTC-1, we still detect these events with nearly similar detection confidence levels in stage 2 but with a computational gain in the matched filtering by a factor of $\sim20$. A comparison of the recovered events' FARs, network SNRs ($\hat{\rho}_{T}\equiv\sqrt{\hat{\rho}_{H}^2+ \hat{\rho}_{L}^2}$), and redshifted chirp mass from the flat search in GWTC-1 and both the stages of the hierarchical search, is given in Table~\ref{table:table2}. 

In our analysis, we recover the loudest events--- GW150914, GW151226, GW170104, GW170608, GW170814, GW170817, and GW170823, with comparable FARs in both the stages of the hierarchical search. However, the network SNRs of these events improve in stage 2. The remaining events--- GW151012, GW170729, and GW170809, see improvements in their FARs and network SNRs in the stage-2 search. 

\section{\label{sec:comparison}Comparison with the flat search}
%

In this section, we compare the search sensitivities of hierarchical and flat search pipelines using similar detection statistics as defined in Sec.~\ref{sec:ranking}. We also highlight the computational advantages of using the former pipeline over the latter. 
%

\subsection{\label{sec:vtcomp}Comparison of sensitivities}

\begin{table}[ht!]
\centering
\caption{Table depicting the ranges for redshifted component masses, total mass, and dimensionless effective spins for each compact object of injected BBH, BNS, and NSBH sources. }
\begin{tabular}{l l l l}
\hline
Parameter \quad \quad & BBH \quad \quad \quad & BNS \quad \quad \quad & NSBH\\
\hline
$m_{1} (M_\odot)$ & 2.5--150 & 1--2.5 & 2.5--97.5\\
$m_{2} (M_\odot)$ & 2.5--150 &  1--2.5  & 1--2.5\\
$M_{tot} (M_\odot)$ & 5--300 & 2--5 & 3.5--100\\
$\chi_{1,z}$ &  0--0.998 & 0--0.4 & 0--0.998  \\
$\chi_{2,z}$ &  0--0.998 & 0--0.4 & 0--0.4 \\
\hline
\end{tabular}
\label{table:injectionset}
\end{table}
The sensitivity of a search pipeline is measured in terms of the total number of astrophysical signals detected at a given detection statistics and a fixed FAR threshold. In order to measure this quantity, a population of simulated GW signals is injected into the real data and recovered using the search pipeline. 
For a population of binary mergers, uniformly distributed over comoving volume ($V$), one can compute the sensitive reach of the detectors in terms of the volume covered in the given observable time. Suppose that a binary's merger rate is defined by $\mu_{m}$; then, the number of detection that one can make above a certain FAR threshold in $T_{obs}$ observation time is the product of volume, time, and merger rate $\mu_{m} \langle VT \rangle$~\cite{VTpycbc}. The sensitive volume-time $\langle VT \rangle$ over here is defined as
\begin{equation}\label{eq:vt}
    \langle VT \rangle_{\{\vec{\theta}\}} = T_{obs} \int^{\infty}_{0} p(z|\{\vec{\theta}\})\frac{dV}{dz}\frac{1}{(1+z)} dz,
\end{equation}
where $p(z|\{\vec{\theta}\})$ is the probability of recovering a signal with parameters $\vec{\theta}$ at a redshift $z$. 
For a constant value of $\mu_{m}$, the ratio of $VT$ can be exploited to compare the sensitivities of any two search pipelines~\cite{gwtc1,gareth}. 
%

In our study, we compare the search sensitivities of the hierarchical and flat search pipelines by computing the ratio of their $VT$ for a common injection set.

To calculate $VT$ for each pipeline, we inject quadrupolar GW signals from the nonprecessing BBH, BNS, and NSBH like sources into the data. These signals are generated using waveform models \verb+SpinTaylorT4+~\cite{spintaylor} for BNS and \verb+SEOBNRv4_opt+~\cite{seobnr} for BBH and NSBH systems. To remain agnostic about the binary merger population, we distribute the signals obtained from these models uniformly over the chirp distance between 50 and 400 Mpc. We uniformly distribute the component masses for BNS and distribute the logarithms of component mass of BBH and NSBH injections in the ranges provided in Table~\ref{table:injectionset}. Thus, we generate 12,203 BNS and 16,271 BBH and NSBH injections each. 

We inject the generated signals in 5 days of coincident data in O1 observed from September 12, 2015, to September 26, 2015, and analyze it using the flat and hierarchical search pipelines separately. The matched-filtering and coincidence studies in the hierarchical search are carried out as per Sec. ~\ref{sec:pipeline}. In the case of flat search, we perform matched filtering over data segments sampled at $2048$ Hz and identify triggers with $\rho$ and $\tilde{\rho}$ above 4 in each detector. We run a coincidence test over the collected single-detector triggers with the appropriate clustering in time as defined in Sec.~\ref{sec:matchedfilter}. Here, triggers observed within 100 ms of a time window in two detectors are identified and ranked according to their statistic values (see Sec.~\ref{sec:ranking}).

The foreground candidates obtained in both the searches are assigned FARs based on their respective noise backgrounds using similar ranking statistics described in Sec.~\ref{sec:ranking}. In the flat search, we estimate the background by time-sliding triggers across the detectors. Each trigger is shifted by 100 ms in time, and then again, the statistic is estimated. A time slide of 100 ms can generate a large number of triggers. Therefore, we first cluster the candidates within a time window of 10 s and then apply decimation to the background as performed in the flat search. 
In the case of hierarchical search, we assign FARs to the detected candidates after scaling the stage-1 background, as described in Sec. \ref{sec:significance}. The recovered candidates via clustering over statistic values are then sorted with respect to their FARs. A highly ranked candidate with a FAR value below 1 per year~\cite{gwtc1} and falling within 1 s of merger time is marked as a detected injection in both the searches. 
\begin{figure}[ht]
    \centering
    \includegraphics[scale=.24]{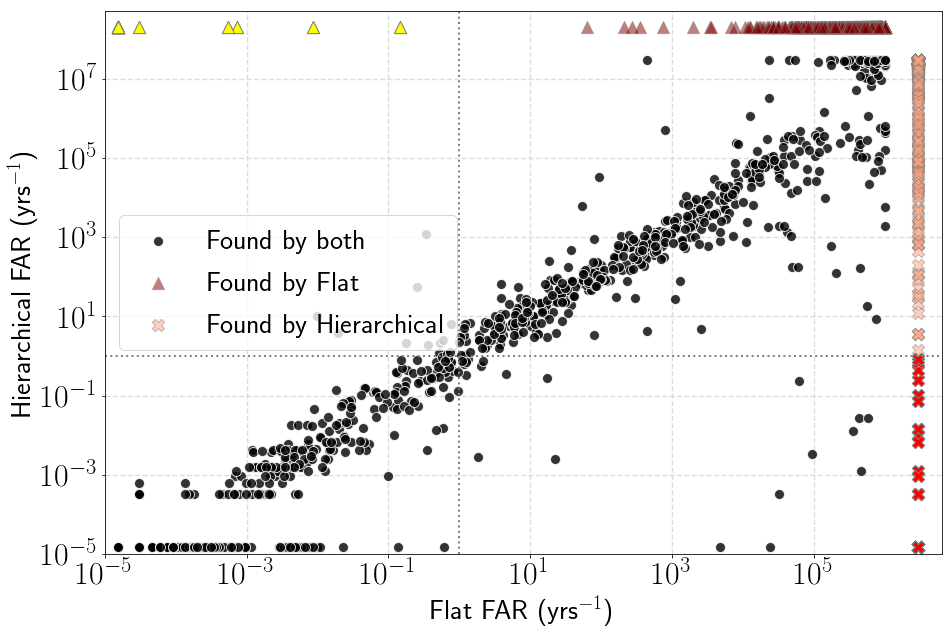}
    \includegraphics[scale=.24]{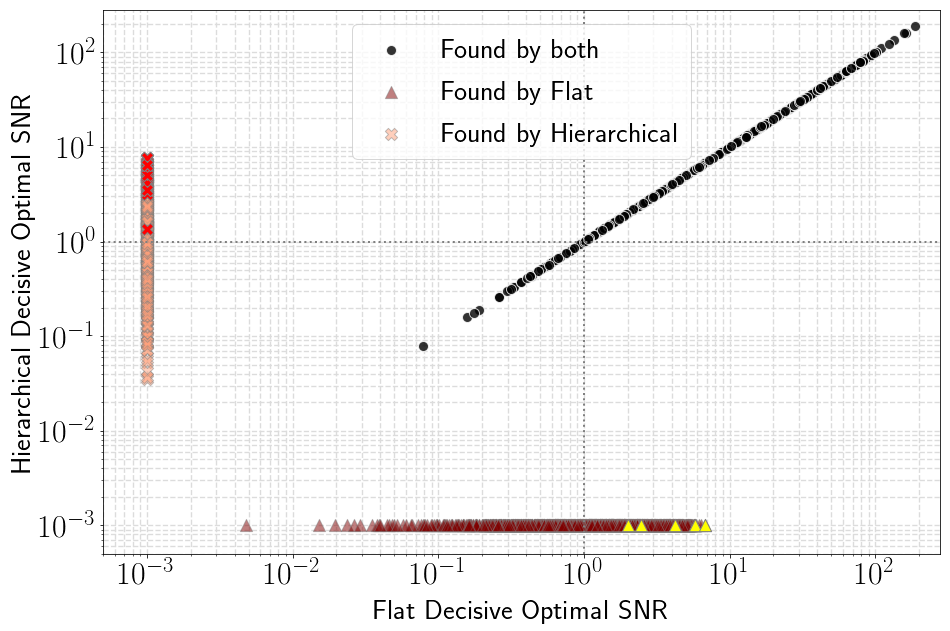}
    \caption{Scatter plots of FARs (top) and decisive optimal SNRs (bottom) for the injections recovered in the hierarchical search vs. flat search. The injections found by both searches are represented by circular points (black). The other markers, cross and triangle, show the injections found by only one search. A few of these points for the flat (yellow) and hierarchical (red) search show low FARs and injected optimal SNRs. The concentration of the points near the diagonal in the top panel implies that the estimated FARs by both the searches are reasonably close. A vertical and horizontal line in the plot shows FAR of 1 per year in the top panel and a decisive optimal SNR of 1 in the bottom panel. The bottom panel confirms that the injections which are not detected by either of the searches were for low ($\sim<8$) decisive optimal SNR.}
    \label{fig:farcomp}
\end{figure} 

 Figure~\ref{fig:farcomp} compares the sensitivities of the hierarchical and flat search. As can be seen in the top panel, most of the injections are recovered with comparable FARs by both searches. We infer this result from the high density of scattered points lying near the diagonal of the plot. We also see that some injections are only recovered by one search. However, these stand-alone recoveries in the majority have a low astrophysical significance. A few of these injections show low FARs, for instance, the injections recovered by only hierarchical search represented in color in Fig.~\ref{fig:farcomp}. A follow-up study on these significant detections showed that these injections were made at very low optimal SNRs (see bottom panel Fig.~\ref{fig:farcomp}) and were likely recovered due to coincidence with noise fluctuation around the injection time. In the other case where injections are recovered by only flat search, hierarchical search misses these injections because stage-1 search fails to recover them. 



\begin{figure*}[ht]
    \centering
    \includegraphics[scale=.51]{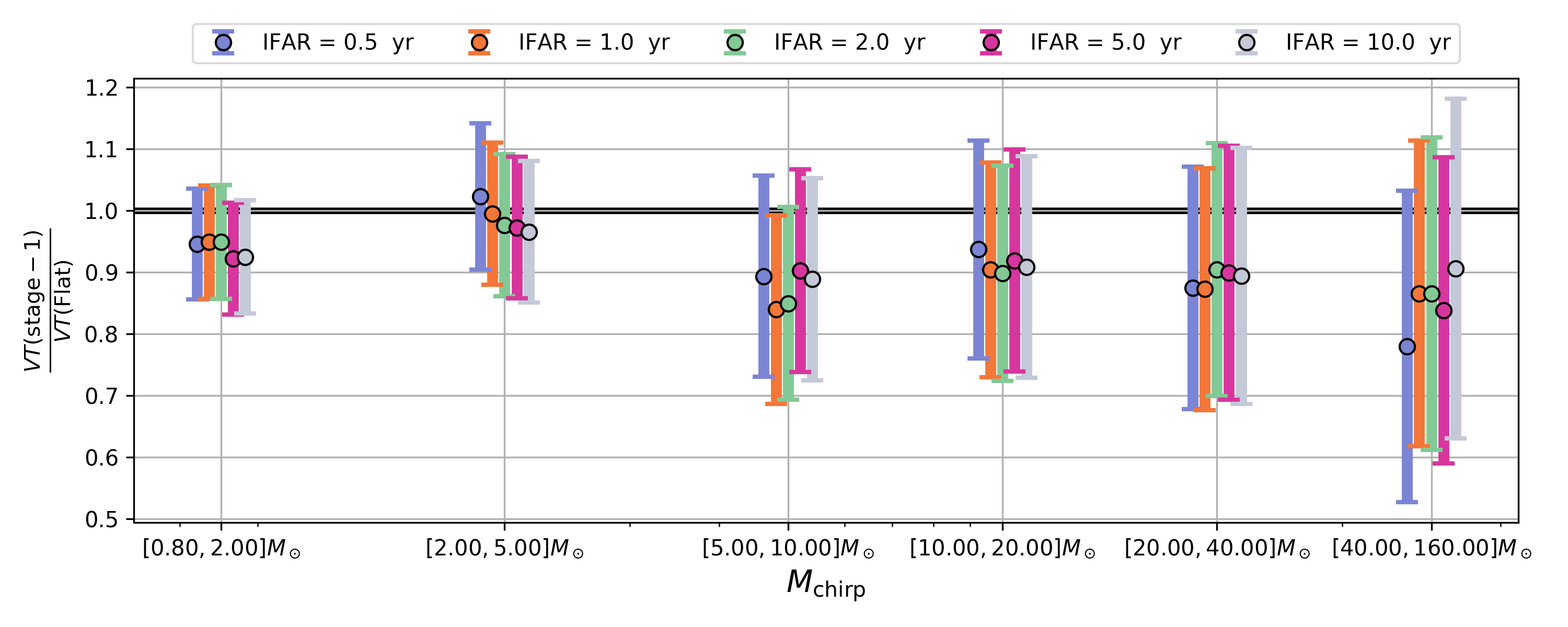}
    \includegraphics[scale=.51]{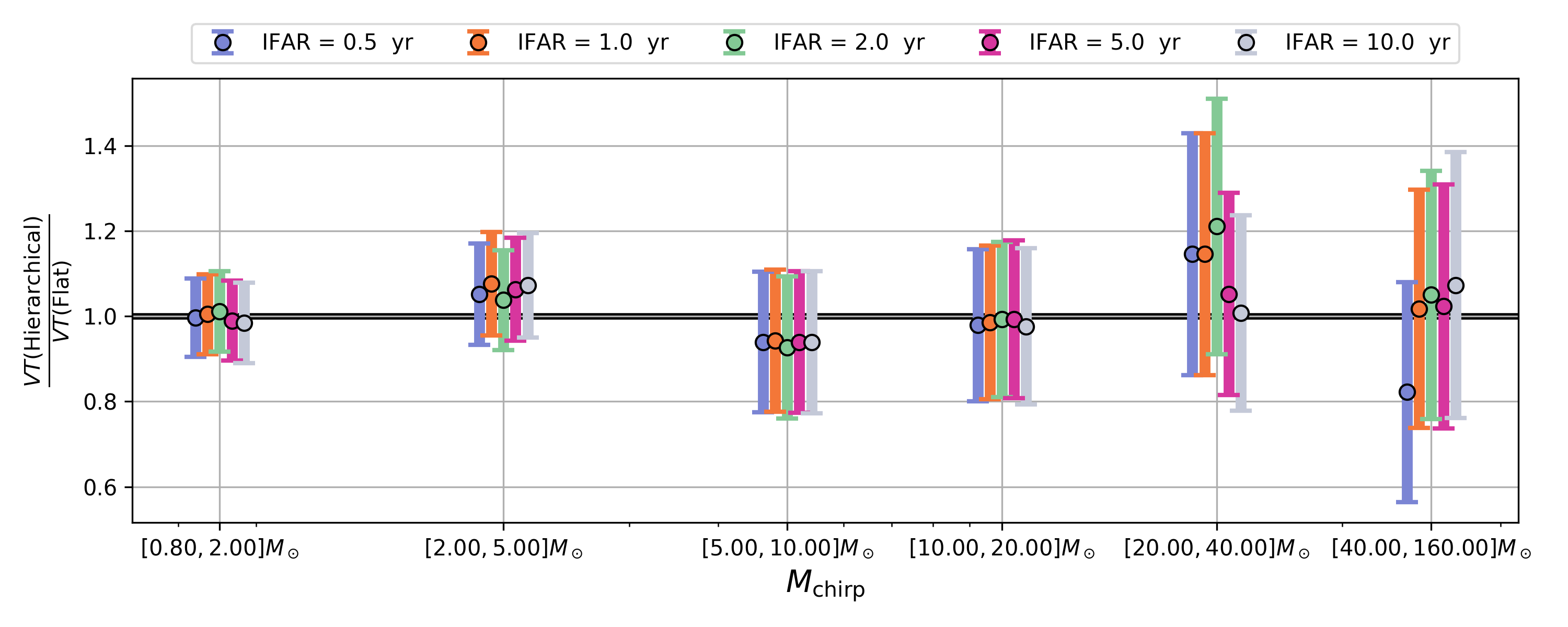}
    \caption{Comparison of volume $\times$ time (VT) ratio of (top) stage 1, and (bottom) stage 2 (or hierarchical) with flat search. The sensitivity of stage-1 search drops for higher chirp mass bins across all IFAR bins in the top panel. In the bottom panel, the VT ratio improves across entire chirp mass and IFAR bins, maintaining the overall sensitivity of hierarchical search comparable to flat.}
    \label{fig:vtplot}
\end{figure*}

The FAR comparison in Fig~\ref{fig:farcomp} shows that both flat and hierarchical search performs similarly for loud CBC injection. However, the sensitivity towards detecting fainter injections varies for both searches. This conclusion is further supported by the $VT$ comparison in Fig~\ref{fig:vtplot}. In the top panel of Fig.~\ref{fig:vtplot}, we see that the sensitivity of stage-1 search is lower than flat search across all the chirp mass and IFAR \footnote{Inverse false alarm rate (IFAR = $\frac{1}{\text{FAR}}$)} bins. This result is expected as the loss in matched-filter SNRs is bound to happen in stage 1 due to low sampling rates and the use of a coarse bank. However, performing a stage-2 search on the potential foreground candidates from stage 1 retains the overall sensitivity of the search pipeline, which can be viewed in the bottom panel of Fig.~\ref{fig:vtplot}. In this plot, we see that the sensitivity of hierarchical search remains consistent with the flat search with $VT$ ratio varying between a factor of $1\pm1.042$ and $1\pm0.08$ for IFAR of $10$ y depending on the chirp mass bins.


\subsection{\label{sec:computationalcost}Comparison of computational efficiencies} 
Now we estimate the computational cost of matched filtering for the flat and hierarchical search. 

The computational cost of matched filtering relies on the number of FFT operations performed on a segment using a bank of templates. As defined previously, FFT operations scale as $N \log_{2} N$. In the case of flat search, we filter a data segment sampled at $2048$ Hz with the entire flat bank. If the segment is of length $512$ s, then $N$ in the flat search is $512\times2048$, and the number of matched-filter operations is $~512\times2048\times428,725\times~\log_{2}(512\times2048)$, where $428,725$ represents the number of templates in the flat bank. 
%
\begin{figure}[ht!]
    \centering
    \includegraphics[scale=.35]{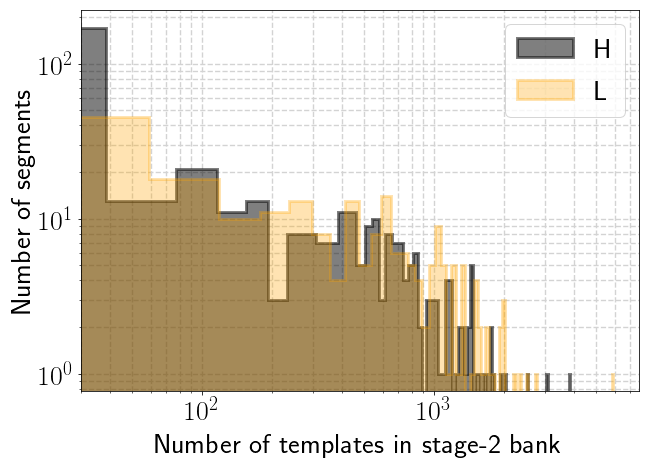}
    \caption{Histogram depicting the number of templates in stage-2 bank generated for each data segment.}
    \label{fig:comp}
\end{figure} 

In the case of hierarchical search, the total number of matched-filter operations is the sum of the number of FFTs performed in stage 1 and stage 2. Since in stage-1 search we matched filter a data segment sampled at $512$ Hz using the coarse bank containing $85,080$ templates, the number of matched-filter operations becomes $512\times512\times85,080~\log_{2}(512\times512)$. If the same segment gets followed up to the stage-2 search, the number of matched-filter operations reduces due to fewer templates in a stage-2 bank. The number of templates in this bank can vary for each segment and detector, as can be seen from Fig.~\ref{fig:comp}. Thus, we compute the total number of FFT operations for all the flat and combined stages of the hierarchical search segments. 
To estimate the overall gain in the computational speed, we take the ratio of the computed FFT operations for the flat to the hierarchical search. 

We first define the following quantities:
\begin{table}[ht!]
\centering
\begin{tabular}{ll}
\hline
 Quantity  &  \quad \quad \quad \quad \quad \quad \quad Description\\
\hline
$\Nsg$ & Total number of data segments in two detectors \\
$\tsg$ & Duration of each segment \\
$\ffl$ & Sampling rate for flat and stage-2 search \\ 
$\fc$ & Sampling rate for stage 1 \\
$\Nt$ & Number of templates in the flat bank \\
$\Ns$ & Number of templates in the stage-1 bank \\
$\Nss$ & Total number of templates for {\it all} the segments\\
~ & used in the stage-2 search\\
\hline
\end{tabular}
\end{table}

Let,
\bea
 \Ofl &=& k~\ffl \tsg \log_2 (\ffl \tsg) \,, \no \\
 \Oc &=& k~\fc \tsg \log_2 (\fc \tsg) \,,
\eea
where $\Ofl$ and $\Oc$ are the number of floating-point operations required to perform a FFT for a segment at the flat and coarse sampling rates, respectively. $k$ is a factor of few which cancels out from the numerator and denominator. Thus, the gain is given by
 
 \begin{equation}
{\rm gain} \approx  \frac{\Nsg ~\Nt ~ \Ofl}{\Nsg \Ns \Oc + \Nss \Ofl} \,.
\end{equation}

While the number of templates in flat search and stage 1 is fixed for all the segments, it varies for each segment in stage 2 as only specific triggers are followed up and filtered using a stage-2 bank. The total area of the histograms for the two detectors together shown in Fig.~\ref{fig:comp} provides us with $\Nss$. Since $\Nss$ is much smaller than $\Nsg ~ \Ns$, the computation in stage 1 dominates the cost, so the stage-2 cost does not affect the gain.

Substituting the numerical values, $\Nsg = 390~[H]+225~[L]=615,~\tsg = 512 {\rm~s},~\ffl = 2048~{\rm~Hz},~\fc = 512~{\rm Hz},~\Nt = 428,725,~\Ns = 85,080,$ and $\Nss = (132,036 ~[H]+132,134~[L])=264,170$, the gain yielded is 22 for the analysis. We do not expect this number to change significantly for different observing runs.
We also compare the actual CPU core hours used by the flat and hierarchical search for performing the matched-filtering operations. We found that the total CPU core hours used by the hierarchical search are around 824.16 and 547.37, respectively, for the Hanford and Livingston detectors. These numbers are nearly 1/19 times the number obtained for the flat search, i.e., 15,471.81 for Hanford and 10,478.64 for Livingston. 
Thus, we conclude that with the present setting, {\em the hierarchical search provides an overall computational speed-up by a factor of $~\sim20$}.
%

\section{\label{sec:conclusion}Conclusion and discussion}

Efficient searches for GWs originating from CBCs can expand the size and dimensionality of the search parameter space to detect interesting sources with present and future detectors. The hierarchical search is perhaps the most straightforward approach that brings more than one order of magnitude enhancement in the efficiency without compromising the robustness of the search. In this work, we successfully demonstrate the efficiency of hierarchical search by applying the analysis on the first two observing runs of Advanced LIGO. By introducing essential modifications to the previously developed codes, we transform them into a complete analysis pipeline~\cite{kanchan}. We improve the selection criteria for single-detector triggers using chi-square and sine-Gaussian vetoes to reweigh matched-filter SNRs. We also implement coincident detection statistics formulated in~\cite{nitzphasetd,gareth} in the hierarchical search that utilizes phase and time differences between detectors and detection parameters, significantly reducing false alarms due to noise events. With our pipeline, we recover all the events in the LIGO-Virgo Collaboration's official transient catalog, GWTC-1, detected by the standard PyCBC analysis with nearly the same statistical confidence and a whopping factor of 20 computational speed-up. This work also demonstrates that hierarchical search is at hand for production analysis of the present and upcoming datasets from ground-based detectors.


Following~\citet{bug}, we estimate the detected candidates' significance by scaling the noise background obtained in stage 1 with a factor close to the speed-up factor. 
Although the argument on assigning significance to detected candidates using this background may not be so rigorous, our work shows that this prescription works. The background estimation for the hierarchical search needs more scrutiny, and our future goal is to address this issue. It is outside the scope of the present investigation because an in-depth mathematical and statistical analysis of the empirical background estimation using time slides will be required. 
While the outcome of this exercise builds enough confidence for application in production runs that are otherwise restrictive due to computational cost, we plan to carry out an extensive study focused on accurate background estimation for the hierarchical search.

In our opinion, the hierarchical search pipeline can be used for ambitious searches that are currently deferred due to computational limitations. For instance, a search for binaries with nonaligned spins and subsolar sources requires an enormous number of templates. With hierarchical search, we can attempt to carry out their search at feasible computation cost without compromising the accuracy of sensitivity of the search. The hierarchical strategy could also reduce the computation cost of low-latency searches, which we plan to demonstrate in the future. 
Developing a comprehensive offline or a low-latency search for such sources is an arduous task ahead, and more sophisticated techniques will have to be brought in, in the coming years. Nevertheless, the hierarchical search is a major step in this direction that should be exploited.


\begin{acknowledgments}
The authors acknowledge the computational resources provided by the IUCAA LDG cluster Sarathi, LIGO Laboratory, and are supported by National Science Foundation Grants. The authors acknowledge support from Soumen Roy in providing code and helping in the generation of banks. The authors are grateful for the valuable discussions from Shreejit P. Jadhav at various stages of this work. K. S. acknowledges technical support for cluster-related issues from Deepak Bankar. The hierarchical search pipeline uses PyCBC version 1.16.13 and is built upon LALSuite~\cite{lalsuite}, NumPy~\cite{numpy}, SciPy~\cite{scipy}, and Astropy~\cite{astropy}. K. S. acknowledges the Inter-University Centre of Astronomy and Astrophysics (IUCAA), India, for the fellowship support. B. G. acknowledges the support of the Max Planck Society. S. M. acknowledges support from the Department of Science and Technology (DST), Ministry of Science and Technology, India, provided under the Swarna Jayanti Fellowships scheme. S. V. D. acknowledges the support of the Senior Scientist Platinum Jubilee Fellowship from NASI, India. This manuscript has been assigned a LIGO Document No. LIGO-P2100202.
\end{acknowledgments}



\nocite{*}

\bibliography{apssamp}

\providecommand{\noopsort}[1]{}\providecommand{\singleletter}[1]{#1}%
\begin{thebibliography}{66}%
\makeatletter
\providecommand \@ifxundefined [1]{%
 \@ifx{#1\undefined}
}%
\providecommand \@ifnum [1]{%
 \ifnum #1\expandafter \@firstoftwo
 \else \expandafter \@secondoftwo
 \fi
}%
\providecommand \@ifx [1]{%
 \ifx #1\expandafter \@firstoftwo
 \else \expandafter \@secondoftwo
 \fi
}%
\providecommand \natexlab [1]{#1}%
\providecommand \enquote  [1]{``#1''}%
\providecommand \bibnamefont  [1]{#1}%
\providecommand \bibfnamefont [1]{#1}%
\providecommand \citenamefont [1]{#1}%
\providecommand \href@noop [0]{\@secondoftwo}%
\providecommand \href [0]{\begingroup \@sanitize@url \@href}%
\providecommand \@href[1]{\@@startlink{#1}\@@href}%
\providecommand \@@href[1]{\endgroup#1\@@endlink}%
\providecommand \@sanitize@url [0]{\catcode `\\12\catcode `\$12\catcode
  `\&12\catcode `\#12\catcode `\^12\catcode `\_12\catcode `\%12\relax}%
\providecommand \@@startlink[1]{}%
\providecommand \@@endlink[0]{}%
\providecommand \url  [0]{\begingroup\@sanitize@url \@url }%
\providecommand \@url [1]{\endgroup\@href {#1}{\urlprefix }}%
\providecommand \urlprefix  [0]{URL }%
\providecommand \Eprint [0]{\href }%
\providecommand \doibase [0]{https://doi.org/}%
\providecommand \selectlanguage [0]{\@gobble}%
\providecommand \bibinfo  [0]{\@secondoftwo}%
\providecommand \bibfield  [0]{\@secondoftwo}%
\providecommand \translation [1]{[#1]}%
\providecommand \BibitemOpen [0]{}%
\providecommand \bibitemStop [0]{}%
\providecommand \bibitemNoStop [0]{.\EOS\space}%
\providecommand \EOS [0]{\spacefactor3000\relax}%
\providecommand \BibitemShut  [1]{\csname bibitem#1\endcsname}%
\let\auto@bib@innerbib\@empty
\bibitem [{\citenamefont {Abbott}\ \emph
  {et~al.}(2016{\natexlab{a}})\citenamefont {Abbott} \emph
  {et~al.}}]{firstevent}%
  \BibitemOpen
  \bibfield  {author} {\bibinfo {author} {\bibfnamefont {B.~P.}\ \bibnamefont
  {Abbott}} \emph {et~al.},\ }\href
  {https://doi.org/10.1103/PhysRevLett.116.061102} {\bibfield  {journal}
  {\bibinfo  {journal} {Phys. Rev. Lett.}\ }\textbf {\bibinfo {volume} {116}},\
  \bibinfo {pages} {061102} (\bibinfo {year} {2016}{\natexlab{a}})}\BibitemShut
  {NoStop}%
\bibitem [{\citenamefont {Abbott}\ \emph
  {et~al.}(2016{\natexlab{b}})\citenamefont {Abbott} \emph
  {et~al.}}]{advligo1}%
  \BibitemOpen
  \bibfield  {author} {\bibinfo {author} {\bibfnamefont {B.~P.}\ \bibnamefont
  {Abbott}} \emph {et~al.},\ }\href
  {https://doi.org/10.1103/PhysRevLett.116.131103} {\bibfield  {journal}
  {\bibinfo  {journal} {Phys. Rev. Lett.}\ }\textbf {\bibinfo {volume} {116}},\
  \bibinfo {pages} {131103} (\bibinfo {year} {2016}{\natexlab{b}})}\BibitemShut
  {NoStop}%
\bibitem [{\citenamefont {Abbott}\ \emph
  {et~al.}(2016{\natexlab{c}})\citenamefont {Abbott} \emph
  {et~al.}}]{advligo2}%
  \BibitemOpen
  \bibfield  {author} {\bibinfo {author} {\bibfnamefont {B.~P.}\ \bibnamefont
  {Abbott}} \emph {et~al.},\ }\href
  {https://doi.org/10.1103/PhysRevD.93.112004} {\bibfield  {journal} {\bibinfo
  {journal} {Phys. Rev. D}\ }\textbf {\bibinfo {volume} {93}},\ \bibinfo
  {pages} {112004} (\bibinfo {year} {2016}{\natexlab{c}})}\BibitemShut
  {NoStop}%
\bibitem [{\citenamefont {Klimenko}\ \emph {et~al.}(2016)\citenamefont
  {Klimenko}, \citenamefont {Vedovato}, \citenamefont {Drago}, \citenamefont
  {Salemi}, \citenamefont {Tiwari}, \citenamefont {Prodi}, \citenamefont
  {Lazzaro}, \citenamefont {Ackley}, \citenamefont {Tiwari}, \citenamefont
  {Da~Silva},\ and\ \citenamefont {Mitselmakher}}]{cwb}%
  \BibitemOpen
  \bibfield  {author} {\bibinfo {author} {\bibfnamefont {S.}~\bibnamefont
  {Klimenko}}, \bibinfo {author} {\bibfnamefont {G.}~\bibnamefont {Vedovato}},
  \bibinfo {author} {\bibfnamefont {M.}~\bibnamefont {Drago}}, \bibinfo
  {author} {\bibfnamefont {F.}~\bibnamefont {Salemi}}, \bibinfo {author}
  {\bibfnamefont {V.}~\bibnamefont {Tiwari}}, \bibinfo {author} {\bibfnamefont
  {G.~A.}\ \bibnamefont {Prodi}}, \bibinfo {author} {\bibfnamefont
  {C.}~\bibnamefont {Lazzaro}}, \bibinfo {author} {\bibfnamefont
  {K.}~\bibnamefont {Ackley}}, \bibinfo {author} {\bibfnamefont
  {S.}~\bibnamefont {Tiwari}}, \bibinfo {author} {\bibfnamefont {C.~F.}\
  \bibnamefont {Da~Silva}},\ and\ \bibinfo {author} {\bibfnamefont
  {G.}~\bibnamefont {Mitselmakher}},\ }\href
  {https://doi.org/10.1103/PhysRevD.93.042004} {\bibfield  {journal} {\bibinfo
  {journal} {Phys. Rev. D}\ }\textbf {\bibinfo {volume} {93}},\ \bibinfo
  {pages} {042004} (\bibinfo {year} {2016})}\BibitemShut {NoStop}%
\bibitem [{\citenamefont {Messick}\ \emph {et~al.}(2017)\citenamefont
  {Messick}, \citenamefont {Blackburn}, \citenamefont {Brady}, \citenamefont
  {Brockill}, \citenamefont {Cannon}, \citenamefont {Cariou}, \citenamefont
  {Caudill}, \citenamefont {Chamberlin}, \citenamefont {Creighton},
  \citenamefont {Everett} \emph {et~al.}}]{gstlal1}%
  \BibitemOpen
  \bibfield  {author} {\bibinfo {author} {\bibfnamefont {C.}~\bibnamefont
  {Messick}}, \bibinfo {author} {\bibfnamefont {K.}~\bibnamefont {Blackburn}},
  \bibinfo {author} {\bibfnamefont {P.}~\bibnamefont {Brady}}, \bibinfo
  {author} {\bibfnamefont {P.}~\bibnamefont {Brockill}}, \bibinfo {author}
  {\bibfnamefont {K.}~\bibnamefont {Cannon}}, \bibinfo {author} {\bibfnamefont
  {R.}~\bibnamefont {Cariou}}, \bibinfo {author} {\bibfnamefont
  {S.}~\bibnamefont {Caudill}}, \bibinfo {author} {\bibfnamefont {S.~J.}\
  \bibnamefont {Chamberlin}}, \bibinfo {author} {\bibfnamefont {J.~D.~E.}\
  \bibnamefont {Creighton}}, \bibinfo {author} {\bibfnamefont {R.}~\bibnamefont
  {Everett}}, \emph {et~al.},\ }\href
  {https://doi.org/10.1103/PhysRevD.95.042001} {\bibfield  {journal} {\bibinfo
  {journal} {Phys. Rev. D}\ }\textbf {\bibinfo {volume} {95}},\ \bibinfo
  {pages} {042001} (\bibinfo {year} {2017})}\BibitemShut {NoStop}%
\bibitem [{\citenamefont {Usman}\ \emph {et~al.}(2016)\citenamefont {Usman},
  \citenamefont {Nitz}, \citenamefont {Harry}, \citenamefont {Biwer},
  \citenamefont {Brown}, \citenamefont {Cabero}, \citenamefont {Capano},
  \citenamefont {Canton}, \citenamefont {Dent}, \citenamefont {Fairhurst} \emph
  {et~al.}}]{usman}%
  \BibitemOpen
  \bibfield  {author} {\bibinfo {author} {\bibfnamefont {S.~A.}\ \bibnamefont
  {Usman}}, \bibinfo {author} {\bibfnamefont {A.~H.}\ \bibnamefont {Nitz}},
  \bibinfo {author} {\bibfnamefont {I.~W.}\ \bibnamefont {Harry}}, \bibinfo
  {author} {\bibfnamefont {C.~M.}\ \bibnamefont {Biwer}}, \bibinfo {author}
  {\bibfnamefont {D.~A.}\ \bibnamefont {Brown}}, \bibinfo {author}
  {\bibfnamefont {M.}~\bibnamefont {Cabero}}, \bibinfo {author} {\bibfnamefont
  {C.~D.}\ \bibnamefont {Capano}}, \bibinfo {author} {\bibfnamefont {T.~D.}\
  \bibnamefont {Canton}}, \bibinfo {author} {\bibfnamefont {T.}~\bibnamefont
  {Dent}}, \bibinfo {author} {\bibfnamefont {S.}~\bibnamefont {Fairhurst}},
  \emph {et~al.},\ }\href {https://doi.org/10.1088/0264-9381/33/21/215004}
  {\bibfield  {journal} {\bibinfo  {journal} {Classical Quantum Gravity}\
  }\textbf {\bibinfo {volume} {33}},\ \bibinfo {pages} {215004} (\bibinfo
  {year} {2016})}\BibitemShut {NoStop}%
\bibitem [{\citenamefont {Abbott}\ \emph
  {et~al.}(2019{\natexlab{a}})\citenamefont {Abbott} \emph {et~al.}}]{gwtc1}%
  \BibitemOpen
  \bibfield  {author} {\bibinfo {author} {\bibfnamefont {B.~P.}\ \bibnamefont
  {Abbott}} \emph {et~al.},\ }\href {https://doi.org/10.1103/PhysRevX.9.031040}
  {\bibfield  {journal} {\bibinfo  {journal} {Phys. Rev. X}\ }\textbf {\bibinfo
  {volume} {9}},\ \bibinfo {pages} {031040} (\bibinfo {year}
  {2019}{\natexlab{a}})}\BibitemShut {NoStop}%
\bibitem [{\citenamefont {Nitz}\ \emph {et~al.}(2019)\citenamefont {Nitz},
  \citenamefont {Capano}, \citenamefont {Nielsen}, \citenamefont {Reyes},
  \citenamefont {White}, \citenamefont {Brown},\ and\ \citenamefont
  {Krishnan}}]{ogc1}%
  \BibitemOpen
  \bibfield  {author} {\bibinfo {author} {\bibfnamefont {A.~H.}\ \bibnamefont
  {Nitz}}, \bibinfo {author} {\bibfnamefont {C.}~\bibnamefont {Capano}},
  \bibinfo {author} {\bibfnamefont {A.~B.}\ \bibnamefont {Nielsen}}, \bibinfo
  {author} {\bibfnamefont {S.}~\bibnamefont {Reyes}}, \bibinfo {author}
  {\bibfnamefont {R.}~\bibnamefont {White}}, \bibinfo {author} {\bibfnamefont
  {D.~A.}\ \bibnamefont {Brown}},\ and\ \bibinfo {author} {\bibfnamefont
  {B.}~\bibnamefont {Krishnan}},\ }\href
  {https://doi.org/10.3847/1538-4357/ab0108} {\bibfield  {journal} {\bibinfo
  {journal} {Astrophys. J.}\ }\textbf {\bibinfo {volume} {872}},\ \bibinfo
  {pages} {195} (\bibinfo {year} {2019})}\BibitemShut {NoStop}%
\bibitem [{\citenamefont {Nitz}\ \emph {et~al.}(2020)\citenamefont {Nitz},
  \citenamefont {Dent}, \citenamefont {Davies}, \citenamefont {Kumar},
  \citenamefont {Capano}, \citenamefont {Harry}, \citenamefont {Mozzon},
  \citenamefont {Nuttall}, \citenamefont {Lundgren},\ and\ \citenamefont
  {Tápai}}]{ogc2}%
  \BibitemOpen
  \bibfield  {author} {\bibinfo {author} {\bibfnamefont {A.~H.}\ \bibnamefont
  {Nitz}}, \bibinfo {author} {\bibfnamefont {T.}~\bibnamefont {Dent}}, \bibinfo
  {author} {\bibfnamefont {G.~S.}\ \bibnamefont {Davies}}, \bibinfo {author}
  {\bibfnamefont {S.}~\bibnamefont {Kumar}}, \bibinfo {author} {\bibfnamefont
  {C.~D.}\ \bibnamefont {Capano}}, \bibinfo {author} {\bibfnamefont
  {I.}~\bibnamefont {Harry}}, \bibinfo {author} {\bibfnamefont
  {S.}~\bibnamefont {Mozzon}}, \bibinfo {author} {\bibfnamefont
  {L.}~\bibnamefont {Nuttall}}, \bibinfo {author} {\bibfnamefont
  {A.}~\bibnamefont {Lundgren}},\ and\ \bibinfo {author} {\bibfnamefont
  {M.}~\bibnamefont {Tápai}},\ }\href
  {https://doi.org/10.3847/1538-4357/ab733f} {\bibfield  {journal} {\bibinfo
  {journal} {Astrophys. J.}\ }\textbf {\bibinfo {volume} {891}},\ \bibinfo
  {pages} {123} (\bibinfo {year} {2020})}\BibitemShut {NoStop}%
\bibitem [{\citenamefont {Venumadhav}\ \emph {et~al.}(2019)\citenamefont
  {Venumadhav}, \citenamefont {Zackay}, \citenamefont {Roulet}, \citenamefont
  {Dai},\ and\ \citenamefont {Zaldarriaga}}]{venumadhav1}%
  \BibitemOpen
  \bibfield  {author} {\bibinfo {author} {\bibfnamefont {T.}~\bibnamefont
  {Venumadhav}}, \bibinfo {author} {\bibfnamefont {B.}~\bibnamefont {Zackay}},
  \bibinfo {author} {\bibfnamefont {J.}~\bibnamefont {Roulet}}, \bibinfo
  {author} {\bibfnamefont {L.}~\bibnamefont {Dai}},\ and\ \bibinfo {author}
  {\bibfnamefont {M.}~\bibnamefont {Zaldarriaga}},\ }\href
  {https://doi.org/10.1103/PhysRevD.100.023011} {\bibfield  {journal} {\bibinfo
   {journal} {Phys. Rev. D}\ }\textbf {\bibinfo {volume} {100}},\ \bibinfo
  {pages} {023011} (\bibinfo {year} {2019})}\BibitemShut {NoStop}%
\bibitem [{\citenamefont {Aasi}\ \emph {et~al.}(2015)\citenamefont {Aasi},
  \citenamefont {Abbott}, \citenamefont {Abbott}, \citenamefont {Abbott},
  \citenamefont {Abernathy}, \citenamefont {Ackley}, \citenamefont {Adams},
  \citenamefont {T.}, \citenamefont {Addesso} \emph {et~al.}}]{asi}%
  \BibitemOpen
  \bibfield  {author} {\bibinfo {author} {\bibfnamefont {J.}~\bibnamefont
  {Aasi}}, \bibinfo {author} {\bibfnamefont {B.~P.}\ \bibnamefont {Abbott}},
  \bibinfo {author} {\bibfnamefont {R.}~\bibnamefont {Abbott}}, \bibinfo
  {author} {\bibfnamefont {T.}~\bibnamefont {Abbott}}, \bibinfo {author}
  {\bibfnamefont {M.~R.}\ \bibnamefont {Abernathy}}, \bibinfo {author}
  {\bibfnamefont {K.}~\bibnamefont {Ackley}}, \bibinfo {author} {\bibfnamefont
  {C.}~\bibnamefont {Adams}}, \bibinfo {author} {\bibfnamefont
  {A.}~\bibnamefont {T.}}, \bibinfo {author} {\bibfnamefont {P.}~\bibnamefont
  {Addesso}}, \emph {et~al.},\ }\href
  {https://doi.org/10.1088/0264-9381/32/7/074001} {\bibfield  {journal}
  {\bibinfo  {journal} {Classical Quantum Gravity}\ }\textbf {\bibinfo {volume}
  {32}},\ \bibinfo {pages} {074001} (\bibinfo {year} {2015})}\BibitemShut
  {NoStop}%
\bibitem [{\citenamefont {Acernese}\ \emph {et~al.}(2014)\citenamefont
  {Acernese}, \citenamefont {Agathos}, \citenamefont {Agatsuma}, \citenamefont
  {Aisa}, \citenamefont {Allemandou}, \citenamefont {Allocca}, \citenamefont
  {Amarni}, \citenamefont {Astone}, \citenamefont {Balestri}, \citenamefont
  {Ballardin} \emph {et~al.}}]{advancedvirgo}%
  \BibitemOpen
  \bibfield  {author} {\bibinfo {author} {\bibfnamefont {F.}~\bibnamefont
  {Acernese}}, \bibinfo {author} {\bibfnamefont {M.}~\bibnamefont {Agathos}},
  \bibinfo {author} {\bibfnamefont {K.}~\bibnamefont {Agatsuma}}, \bibinfo
  {author} {\bibfnamefont {D.}~\bibnamefont {Aisa}}, \bibinfo {author}
  {\bibfnamefont {N.}~\bibnamefont {Allemandou}}, \bibinfo {author}
  {\bibfnamefont {A.}~\bibnamefont {Allocca}}, \bibinfo {author} {\bibfnamefont
  {J.}~\bibnamefont {Amarni}}, \bibinfo {author} {\bibfnamefont
  {P.}~\bibnamefont {Astone}}, \bibinfo {author} {\bibfnamefont
  {G.}~\bibnamefont {Balestri}}, \bibinfo {author} {\bibfnamefont
  {G.}~\bibnamefont {Ballardin}}, \emph {et~al.},\ }\href
  {https://doi.org/10.1088/0264-9381/32/2/024001} {\bibfield  {journal}
  {\bibinfo  {journal} {Classical Quantum Gravity}\ }\textbf {\bibinfo {volume}
  {32}},\ \bibinfo {pages} {024001} (\bibinfo {year} {2014})}\BibitemShut
  {NoStop}%
\bibitem [{\citenamefont {Abbott}\ \emph
  {et~al.}(2021{\natexlab{a}})\citenamefont {Abbott} \emph {et~al.}}]{gwtc2}%
  \BibitemOpen
  \bibfield  {author} {\bibinfo {author} {\bibfnamefont {R.}~\bibnamefont
  {Abbott}} \emph {et~al.},\ }\href
  {https://doi.org/10.1103/PhysRevX.11.021053} {\bibfield  {journal} {\bibinfo
  {journal} {Phys. Rev. X}\ }\textbf {\bibinfo {volume} {11}},\ \bibinfo
  {pages} {021053} (\bibinfo {year} {2021}{\natexlab{a}})}\BibitemShut
  {NoStop}%
\bibitem [{\citenamefont {Abbott}\ \emph
  {et~al.}(2021{\natexlab{b}})\citenamefont {Abbott} \emph {et~al.}}]{gwtc2.1}%
  \BibitemOpen
  \bibfield  {author} {\bibinfo {author} {\bibfnamefont {R.}~\bibnamefont
  {Abbott}} \emph {et~al.},\ }\href {https://arxiv.org/abs/2108.01045} {\
  (\bibinfo {year} {2021}{\natexlab{b}})},\ \Eprint
  {https://arxiv.org/abs/2108.01045} {arXiv:2108.01045} \BibitemShut {NoStop}%
\bibitem [{\citenamefont {Abbott}\ \emph
  {et~al.}(2021{\natexlab{c}})\citenamefont {Abbott} \emph {et~al.}}]{gwtc3}%
  \BibitemOpen
  \bibfield  {author} {\bibinfo {author} {\bibfnamefont {R.}~\bibnamefont
  {Abbott}} \emph {et~al.},\ }\href {https://arxiv.org/abs/2111.03606} {\
  (\bibinfo {year} {2021}{\natexlab{c}})},\ \Eprint
  {https://arxiv.org/abs/2111.03606} {arXiv:2111.03606} \BibitemShut {NoStop}%
\bibitem [{\citenamefont {Abbott}\ \emph
  {et~al.}(2020{\natexlab{a}})\citenamefont {Abbott} \emph
  {et~al.}}]{GW190425}%
  \BibitemOpen
  \bibfield  {author} {\bibinfo {author} {\bibfnamefont {B.~P.}\ \bibnamefont
  {Abbott}} \emph {et~al.},\ }\href {https://doi.org/10.3847/2041-8213/ab75f5}
  {\bibfield  {journal} {\bibinfo  {journal} {Astrophys. J. Lett.}\ }\textbf
  {\bibinfo {volume} {892}},\ \bibinfo {pages} {L3} (\bibinfo {year}
  {2020}{\natexlab{a}})}\BibitemShut {NoStop}%
\bibitem [{\citenamefont {Abbott}\ \emph
  {et~al.}(2020{\natexlab{b}})\citenamefont {Abbott} \emph
  {et~al.}}]{GW190412}%
  \BibitemOpen
  \bibfield  {author} {\bibinfo {author} {\bibfnamefont {B.~P.}\ \bibnamefont
  {Abbott}} \emph {et~al.},\ }\href
  {https://doi.org/10.1103/PhysRevD.102.043015} {\bibfield  {journal} {\bibinfo
   {journal} {Phys. Rev. D}\ }\textbf {\bibinfo {volume} {102}},\ \bibinfo
  {pages} {043015} (\bibinfo {year} {2020}{\natexlab{b}})}\BibitemShut
  {NoStop}%
\bibitem [{\citenamefont {Abbott}\ \emph
  {et~al.}(2020{\natexlab{c}})\citenamefont {Abbott} \emph
  {et~al.}}]{GW190814}%
  \BibitemOpen
  \bibfield  {author} {\bibinfo {author} {\bibfnamefont {R.}~\bibnamefont
  {Abbott}} \emph {et~al.},\ }\href {https://doi.org/10.3847/2041-8213/ab960f}
  {\bibfield  {journal} {\bibinfo  {journal} {Astrophys. J. Lett.}\ }\textbf
  {\bibinfo {volume} {896}},\ \bibinfo {pages} {L44} (\bibinfo {year}
  {2020}{\natexlab{c}})}\BibitemShut {NoStop}%
\bibitem [{\citenamefont {Abbott}\ \emph
  {et~al.}(2020{\natexlab{d}})\citenamefont {Abbott} \emph {et~al.}}]{IMBH}%
  \BibitemOpen
  \bibfield  {author} {\bibinfo {author} {\bibfnamefont {R.}~\bibnamefont
  {Abbott}} \emph {et~al.},\ }\href {https://doi.org/10.3847/2041-8213/aba493}
  {\bibfield  {journal} {\bibinfo  {journal} {Astrophys. J. Lett.}\ }\textbf
  {\bibinfo {volume} {900}},\ \bibinfo {pages} {L13} (\bibinfo {year}
  {2020}{\natexlab{d}})}\BibitemShut {NoStop}%
\bibitem [{\citenamefont {Sathyaprakash}\ and\ \citenamefont
  {Dhurandhar}(1991)}]{svd}%
  \BibitemOpen
  \bibfield  {author} {\bibinfo {author} {\bibfnamefont {B.~S.}\ \bibnamefont
  {Sathyaprakash}}\ and\ \bibinfo {author} {\bibfnamefont {S.~V.}\ \bibnamefont
  {Dhurandhar}},\ }\href {https://doi.org/10.1103/PhysRevD.44.3819} {\bibfield
  {journal} {\bibinfo  {journal} {Phys. Rev. D}\ }\textbf {\bibinfo {volume}
  {44}},\ \bibinfo {pages} {3819} (\bibinfo {year} {1991})}\BibitemShut
  {NoStop}%
\bibitem [{\citenamefont {Dhurandhar}\ and\ \citenamefont
  {Sathyaprakash}(1994)}]{svd-satya}%
  \BibitemOpen
  \bibfield  {author} {\bibinfo {author} {\bibfnamefont {S.~V.}\ \bibnamefont
  {Dhurandhar}}\ and\ \bibinfo {author} {\bibfnamefont {B.~S.}\ \bibnamefont
  {Sathyaprakash}},\ }\href {https://doi.org/10.1103/PhysRevD.49.1707}
  {\bibfield  {journal} {\bibinfo  {journal} {Phys. Rev. D}\ }\textbf {\bibinfo
  {volume} {49}},\ \bibinfo {pages} {1707} (\bibinfo {year}
  {1994})}\BibitemShut {NoStop}%
\bibitem [{\citenamefont {Dhurandhar}\ and\ \citenamefont
  {Schutz}(1994)}]{svd-schutz}%
  \BibitemOpen
  \bibfield  {author} {\bibinfo {author} {\bibfnamefont {S.~V.}\ \bibnamefont
  {Dhurandhar}}\ and\ \bibinfo {author} {\bibfnamefont {B.~F.}\ \bibnamefont
  {Schutz}},\ }\href {https://doi.org/10.1103/PhysRevD.50.2390} {\bibfield
  {journal} {\bibinfo  {journal} {Phys. Rev. D}\ }\textbf {\bibinfo {volume}
  {50}},\ \bibinfo {pages} {2390} (\bibinfo {year} {1994})}\BibitemShut
  {NoStop}%
\bibitem [{\citenamefont {Owen}\ and\ \citenamefont
  {Sathyaprakash}(1999)}]{sathya-owen}%
  \BibitemOpen
  \bibfield  {author} {\bibinfo {author} {\bibfnamefont {B.~J.}\ \bibnamefont
  {Owen}}\ and\ \bibinfo {author} {\bibfnamefont {B.~S.}\ \bibnamefont
  {Sathyaprakash}},\ }\href {https://doi.org/10.1103/PhysRevD.60.022002}
  {\bibfield  {journal} {\bibinfo  {journal} {Phys. Rev. D}\ }\textbf {\bibinfo
  {volume} {60}},\ \bibinfo {pages} {022002} (\bibinfo {year}
  {1999})}\BibitemShut {NoStop}%
\bibitem [{\citenamefont {Allen}\ \emph {et~al.}(2012)\citenamefont {Allen},
  \citenamefont {Anderson}, \citenamefont {Brady}, \citenamefont {Brown},\ and\
  \citenamefont {Creighton}}]{findchirp}%
  \BibitemOpen
  \bibfield  {author} {\bibinfo {author} {\bibfnamefont {B.}~\bibnamefont
  {Allen}}, \bibinfo {author} {\bibfnamefont {W.~G.}\ \bibnamefont {Anderson}},
  \bibinfo {author} {\bibfnamefont {P.~R.}\ \bibnamefont {Brady}}, \bibinfo
  {author} {\bibfnamefont {D.~A.}\ \bibnamefont {Brown}},\ and\ \bibinfo
  {author} {\bibfnamefont {J.~D.~E.}\ \bibnamefont {Creighton}},\ }\href
  {https://doi.org/10.1103/PhysRevD.85.122006} {\bibfield  {journal} {\bibinfo
  {journal} {Phys. Rev. D}\ }\textbf {\bibinfo {volume} {85}},\ \bibinfo
  {pages} {122006} (\bibinfo {year} {2012})}\BibitemShut {NoStop}%
\bibitem [{\citenamefont {Sachdev}\ \emph {et~al.}(2019)\citenamefont
  {Sachdev}, \citenamefont {Caudill}, \citenamefont {Fong}, \citenamefont {Lo},
  \citenamefont {Messick}, \citenamefont {Mukherjee}, \citenamefont {Magee},
  \citenamefont {Tsukad} \emph {et~al.}}]{gstlal2}%
  \BibitemOpen
  \bibfield  {author} {\bibinfo {author} {\bibfnamefont {S.}~\bibnamefont
  {Sachdev}}, \bibinfo {author} {\bibfnamefont {S.}~\bibnamefont {Caudill}},
  \bibinfo {author} {\bibfnamefont {H.}~\bibnamefont {Fong}}, \bibinfo {author}
  {\bibfnamefont {K.~L.~R.}\ \bibnamefont {Lo}}, \bibinfo {author}
  {\bibfnamefont {C.}~\bibnamefont {Messick}}, \bibinfo {author} {\bibfnamefont
  {D.}~\bibnamefont {Mukherjee}}, \bibinfo {author} {\bibfnamefont
  {R.}~\bibnamefont {Magee}}, \bibinfo {author} {\bibfnamefont
  {L.}~\bibnamefont {Tsukad}}, \emph {et~al.},\ }\href
  {https://arxiv.org/abs/1901.08580} {} (\bibinfo {year} {2019}),\ \Eprint
  {https://arxiv.org/abs/1901.08580} {arXiv:1901.08580} \BibitemShut {NoStop}%
\bibitem [{\citenamefont {Hanna}\ \emph {et~al.}(2020)\citenamefont {Hanna},
  \citenamefont {Caudill}, \citenamefont {Messick}, \citenamefont {Reza},
  \citenamefont {Sachdev}, \citenamefont {Tsukada}, \citenamefont {Cannon},
  \citenamefont {Blackburn}, \citenamefont {Creighton}, \citenamefont {Fong}
  \emph {et~al.}}]{gstlal3}%
  \BibitemOpen
  \bibfield  {author} {\bibinfo {author} {\bibfnamefont {C.}~\bibnamefont
  {Hanna}}, \bibinfo {author} {\bibfnamefont {C.}~\bibnamefont {Caudill}},
  \bibinfo {author} {\bibfnamefont {C.}~\bibnamefont {Messick}}, \bibinfo
  {author} {\bibfnamefont {A.}~\bibnamefont {Reza}}, \bibinfo {author}
  {\bibfnamefont {S.}~\bibnamefont {Sachdev}}, \bibinfo {author} {\bibfnamefont
  {L.}~\bibnamefont {Tsukada}}, \bibinfo {author} {\bibfnamefont
  {K.}~\bibnamefont {Cannon}}, \bibinfo {author} {\bibfnamefont
  {K.}~\bibnamefont {Blackburn}}, \bibinfo {author} {\bibfnamefont {J.~D.~E.}\
  \bibnamefont {Creighton}}, \bibinfo {author} {\bibfnamefont {H.}~\bibnamefont
  {Fong}}, \emph {et~al.},\ }\href
  {https://doi.org/10.1103/PhysRevD.101.022003} {\bibfield  {journal} {\bibinfo
   {journal} {Phys. Rev. D}\ }\textbf {\bibinfo {volume} {101}},\ \bibinfo
  {pages} {022003} (\bibinfo {year} {2020})}\BibitemShut {NoStop}%
\bibitem [{\citenamefont {Adams}\ \emph {et~al.}(2016)\citenamefont {Adams},
  \citenamefont {Buskulic}, \citenamefont {Germain}, \citenamefont {Guidi},
  \citenamefont {Marion}, \citenamefont {Montani}, \citenamefont {Mours},
  \citenamefont {Piergiovanni},\ and\ \citenamefont {Wang}}]{mbta1}%
  \BibitemOpen
  \bibfield  {author} {\bibinfo {author} {\bibfnamefont {T.}~\bibnamefont
  {Adams}}, \bibinfo {author} {\bibfnamefont {D.}~\bibnamefont {Buskulic}},
  \bibinfo {author} {\bibfnamefont {V.}~\bibnamefont {Germain}}, \bibinfo
  {author} {\bibfnamefont {G.~M.}\ \bibnamefont {Guidi}}, \bibinfo {author}
  {\bibfnamefont {F.}~\bibnamefont {Marion}}, \bibinfo {author} {\bibfnamefont
  {M.}~\bibnamefont {Montani}}, \bibinfo {author} {\bibfnamefont
  {B.}~\bibnamefont {Mours}}, \bibinfo {author} {\bibfnamefont
  {F.}~\bibnamefont {Piergiovanni}},\ and\ \bibinfo {author} {\bibfnamefont
  {G.}~\bibnamefont {Wang}},\ }\href
  {https://doi.org/10.1088/0264-9381/33/17/175012} {\bibfield  {journal}
  {\bibinfo  {journal} {Classical Quantum Gravity}\ }\textbf {\bibinfo {volume}
  {33}},\ \bibinfo {pages} {175012} (\bibinfo {year} {2016})}\BibitemShut
  {NoStop}%
\bibitem [{\citenamefont {Aubin}\ \emph {et~al.}(2021)\citenamefont {Aubin},
  \citenamefont {Chierici}, \citenamefont {Estevez}, \citenamefont {Greco},
  \citenamefont {Guidi}, \citenamefont {V.}, \citenamefont {Marion},
  \citenamefont {Mours}, \citenamefont {Nitoglia} \emph {et~al.}}]{mbta2}%
  \BibitemOpen
  \bibfield  {author} {\bibinfo {author} {\bibfnamefont {F.}~\bibnamefont
  {Aubin}, \bibfnamefont {F.~adn~Brighenti}}, \bibinfo {author} {\bibfnamefont
  {R.}~\bibnamefont {Chierici}}, \bibinfo {author} {\bibfnamefont
  {D.}~\bibnamefont {Estevez}}, \bibinfo {author} {\bibfnamefont
  {G.}~\bibnamefont {Greco}}, \bibinfo {author} {\bibfnamefont {G.~M.}\
  \bibnamefont {Guidi}}, \bibinfo {author} {\bibfnamefont {J.}~\bibnamefont
  {V.}}, \bibinfo {author} {\bibfnamefont {F.}~\bibnamefont {Marion}}, \bibinfo
  {author} {\bibfnamefont {B.}~\bibnamefont {Mours}}, \bibinfo {author}
  {\bibfnamefont {E.}~\bibnamefont {Nitoglia}}, \emph {et~al.},\ }\href
  {https://doi.org/10.1088/1361-6382/abe913} {\bibfield  {journal} {\bibinfo
  {journal} {Classical Quantum Gravity}\ }\textbf {\bibinfo {volume} {38}},\
  \bibinfo {pages} {095004} (\bibinfo {year} {2021})}\BibitemShut {NoStop}%
\bibitem [{\citenamefont {Davies}\ \emph {et~al.}(2020)\citenamefont {Davies},
  \citenamefont {Dent}, \citenamefont {T\'apai}, \citenamefont {Harry},
  \citenamefont {McIsaac},\ and\ \citenamefont {Nitz}}]{gareth}%
  \BibitemOpen
  \bibfield  {author} {\bibinfo {author} {\bibfnamefont {G.~S.}\ \bibnamefont
  {Davies}}, \bibinfo {author} {\bibfnamefont {T.}~\bibnamefont {Dent}},
  \bibinfo {author} {\bibfnamefont {M.}~\bibnamefont {T\'apai}}, \bibinfo
  {author} {\bibfnamefont {I.}~\bibnamefont {Harry}}, \bibinfo {author}
  {\bibfnamefont {C.}~\bibnamefont {McIsaac}},\ and\ \bibinfo {author}
  {\bibfnamefont {A.~H.}\ \bibnamefont {Nitz}},\ }\href
  {https://doi.org/10.1103/PhysRevD.102.022004} {\bibfield  {journal} {\bibinfo
   {journal} {Phys. Rev. D}\ }\textbf {\bibinfo {volume} {102}},\ \bibinfo
  {pages} {022004} (\bibinfo {year} {2020})}\BibitemShut {NoStop}%
\bibitem [{\citenamefont {Chu}(2017)}]{spiir}%
  \BibitemOpen
  \bibfield  {author} {\bibinfo {author} {\bibfnamefont {Q.}~\bibnamefont
  {Chu}},\ }\href
  {https://research-repository.uwa.edu.au/en/publications/low-latency-detection-and-localization-of-gravitational-waves-fro}
  {\bibinfo {title} {\uppercase{P}h.\uppercase{D}. thesis, \uppercase{T}he
  \uppercase{U}niversity of \uppercase{W}estern \uppercase{A}ustralia}}
  (\bibinfo {year} {2017})\BibitemShut {NoStop}%
\bibitem [{\citenamefont {Canton}\ and\ \citenamefont
  {W.}(2017)}]{tito-ian_templatebank}%
  \BibitemOpen
  \bibfield  {author} {\bibinfo {author} {\bibfnamefont {T.~D.}\ \bibnamefont
  {Canton}}\ and\ \bibinfo {author} {\bibfnamefont {H.~I.}\ \bibnamefont
  {W.}},\ }\href {https://arxiv.org/abs/1705.01845} {} (\bibinfo {year}
  {2017}),\ \Eprint {https://arxiv.org/abs/1705.01845} {arXiv:1705.01845}
  \BibitemShut {NoStop}%
\bibitem [{\citenamefont {Harry}\ \emph {et~al.}(2016)\citenamefont {Harry},
  \citenamefont {Privitera}, \citenamefont {Boh\'e},\ and\ \citenamefont
  {Buonanno}}]{ianharry}%
  \BibitemOpen
  \bibfield  {author} {\bibinfo {author} {\bibfnamefont {I.}~\bibnamefont
  {Harry}}, \bibinfo {author} {\bibfnamefont {S.}~\bibnamefont {Privitera}},
  \bibinfo {author} {\bibfnamefont {A.}~\bibnamefont {Boh\'e}},\ and\ \bibinfo
  {author} {\bibfnamefont {A.}~\bibnamefont {Buonanno}},\ }\href
  {https://doi.org/10.1103/PhysRevD.94.024012} {\bibfield  {journal} {\bibinfo
  {journal} {Phys. Rev. D}\ }\textbf {\bibinfo {volume} {94}},\ \bibinfo
  {pages} {024012} (\bibinfo {year} {2016})}\BibitemShut {NoStop}%
\bibitem [{\citenamefont {Abbott}\ \emph
  {et~al.}(2019{\natexlab{b}})\citenamefont {Abbott} \emph
  {et~al.}}]{subsolar_lvk19}%
  \BibitemOpen
  \bibfield  {author} {\bibinfo {author} {\bibfnamefont {B.~P.}\ \bibnamefont
  {Abbott}} \emph {et~al.},\ }\href
  {https://doi.org/10.1103/PhysRevLett.123.161102} {\bibfield  {journal}
  {\bibinfo  {journal} {Phys. Rev. Lett.}\ }\textbf {\bibinfo {volume} {123}},\
  \bibinfo {pages} {161102} (\bibinfo {year} {2019}{\natexlab{b}})}\BibitemShut
  {NoStop}%
\bibitem [{\citenamefont {Abbott}\ \emph
  {et~al.}(2021{\natexlab{d}})\citenamefont {Abbott} \emph
  {et~al.}}]{subsolar_lvk21}%
  \BibitemOpen
  \bibfield  {author} {\bibinfo {author} {\bibfnamefont {R.}~\bibnamefont
  {Abbott}} \emph {et~al.},\ }\href {https://arxiv.org/abs/2109.12197} {\
  (\bibinfo {year} {2021}{\natexlab{d}})},\ \Eprint
  {https://arxiv.org/abs/2109.12197} {arXiv:2109.12197} \BibitemShut {NoStop}%
\bibitem [{\citenamefont {Nitz}\ and\ \citenamefont
  {Wang}(2021)}]{nitz_wang_subsolar}%
  \BibitemOpen
  \bibfield  {author} {\bibinfo {author} {\bibfnamefont {A.~H.}\ \bibnamefont
  {Nitz}}\ and\ \bibinfo {author} {\bibfnamefont {Y.-F.}\ \bibnamefont
  {Wang}},\ }\href {https://doi.org/10.3847/1538-4357/ac01d9} {\bibfield
  {journal} {\bibinfo  {journal} {Astrophys. J.}\ }\textbf {\bibinfo {volume}
  {915}},\ \bibinfo {pages} {54} (\bibinfo {year} {2021})}\BibitemShut
  {NoStop}%
\bibitem [{\citenamefont {Akutsu}\ \emph {et~al.}(2020)\citenamefont {Akutsu},
  \citenamefont {Ando}, \citenamefont {Arai}, \citenamefont {Arai},
  \citenamefont {Araki}, \citenamefont {Araya}, \citenamefont {Aritomi},
  \citenamefont {Aso}, \citenamefont {Bae}, \citenamefont {Bae} \emph
  {et~al.}}]{kagra}%
  \BibitemOpen
  \bibfield  {author} {\bibinfo {author} {\bibfnamefont {T.}~\bibnamefont
  {Akutsu}}, \bibinfo {author} {\bibfnamefont {M.}~\bibnamefont {Ando}},
  \bibinfo {author} {\bibfnamefont {K.}~\bibnamefont {Arai}}, \bibinfo {author}
  {\bibfnamefont {Y.}~\bibnamefont {Arai}}, \bibinfo {author} {\bibfnamefont
  {S.}~\bibnamefont {Araki}}, \bibinfo {author} {\bibfnamefont
  {S.}~\bibnamefont {Araya}}, \bibinfo {author} {\bibfnamefont
  {N.}~\bibnamefont {Aritomi}}, \bibinfo {author} {\bibfnamefont
  {Y.}~\bibnamefont {Aso}}, \bibinfo {author} {\bibfnamefont {S.}~\bibnamefont
  {Bae}}, \bibinfo {author} {\bibfnamefont {Y.}~\bibnamefont {Bae}}, \emph
  {et~al.},\ }\href {https://arxiv.org/abs/2005.05574} {} (\bibinfo {year}
  {2020}),\ \Eprint {https://arxiv.org/abs/2005.05574} {arXiv:2005.05574}
  \BibitemShut {NoStop}%
\bibitem [{\citenamefont {Iyer}\ \emph {et~al.}(2011)\citenamefont {Iyer} \emph
  {et~al.}}]{ligoindia}%
  \BibitemOpen
  \bibfield  {author} {\bibinfo {author} {\bibfnamefont {B.}~\bibnamefont
  {Iyer}} \emph {et~al.},\ }\href {https://dcc.ligo.org/LIGO-M1100296/public}
  {\bibinfo {title} {\uppercase{LIGO-I}ndia \uppercase{T}ech. rep.}} (\bibinfo
  {year} {2011})\BibitemShut {NoStop}%
\bibitem [{\citenamefont {Mohanty}\ and\ \citenamefont
  {Dhurandhar}(1996)}]{mohanty96}%
  \BibitemOpen
  \bibfield  {author} {\bibinfo {author} {\bibfnamefont {S.~D.}\ \bibnamefont
  {Mohanty}}\ and\ \bibinfo {author} {\bibfnamefont {S.~V.}\ \bibnamefont
  {Dhurandhar}},\ }\href {https://doi.org/10.1103/PhysRevD.54.7108} {\bibfield
  {journal} {\bibinfo  {journal} {Phys. Rev. D}\ }\textbf {\bibinfo {volume}
  {54}},\ \bibinfo {pages} {7108} (\bibinfo {year} {1996})}\BibitemShut
  {NoStop}%
\bibitem [{\citenamefont {Mohanty}(1998)}]{mohanty98}%
  \BibitemOpen
  \bibfield  {author} {\bibinfo {author} {\bibfnamefont {S.~D.}\ \bibnamefont
  {Mohanty}},\ }\href {https://doi.org/10.1103/PhysRevD.57.630} {\bibfield
  {journal} {\bibinfo  {journal} {Phys. Rev. D}\ }\textbf {\bibinfo {volume}
  {57}},\ \bibinfo {pages} {630} (\bibinfo {year} {1998})}\BibitemShut
  {NoStop}%
\bibitem [{\citenamefont {Gadre}\ \emph {et~al.}(2019)\citenamefont {Gadre},
  \citenamefont {Mitra},\ and\ \citenamefont {Dhurandhar}}]{bug}%
  \BibitemOpen
  \bibfield  {author} {\bibinfo {author} {\bibfnamefont {B.}~\bibnamefont
  {Gadre}}, \bibinfo {author} {\bibfnamefont {S.}~\bibnamefont {Mitra}},\ and\
  \bibinfo {author} {\bibfnamefont {S.}~\bibnamefont {Dhurandhar}},\ }\href
  {https://doi.org/10.1103/PhysRevD.99.124035} {\bibfield  {journal} {\bibinfo
  {journal} {Phys. Rev. D}\ }\textbf {\bibinfo {volume} {99}},\ \bibinfo
  {pages} {124035} (\bibinfo {year} {2019})}\BibitemShut {NoStop}%
\bibitem [{\citenamefont {Devine}\ \emph {et~al.}(2016)\citenamefont {Devine},
  \citenamefont {Etienne},\ and\ \citenamefont {McWilliams}}]{seobnr}%
  \BibitemOpen
  \bibfield  {author} {\bibinfo {author} {\bibfnamefont {C.}~\bibnamefont
  {Devine}}, \bibinfo {author} {\bibfnamefont {Z.~B.}\ \bibnamefont
  {Etienne}},\ and\ \bibinfo {author} {\bibfnamefont {S.~T.}\ \bibnamefont
  {McWilliams}},\ }\href {https://doi.org/10.1088/0264-9381/33/12/125025}
  {\bibfield  {journal} {\bibinfo  {journal} {Classical Quantum Gravity}\
  }\textbf {\bibinfo {volume} {33}},\ \bibinfo {pages} {125025} (\bibinfo
  {year} {2016})}\BibitemShut {NoStop}%
\bibitem [{\citenamefont {Prix}(2007)}]{spherecov1}%
  \BibitemOpen
  \bibfield  {author} {\bibinfo {author} {\bibfnamefont {R.}~\bibnamefont
  {Prix}},\ }\href {https://doi.org/10.1088/0264-9381/24/19/s11} {\bibfield
  {journal} {\bibinfo  {journal} {Classical Quantum Gravity}\ }\textbf
  {\bibinfo {volume} {24}},\ \bibinfo {pages} {S481} (\bibinfo {year}
  {2007})}\BibitemShut {NoStop}%
\bibitem [{\citenamefont {Harry}\ \emph {et~al.}(2009)\citenamefont {Harry},
  \citenamefont {Allen},\ and\ \citenamefont {Sathyaprakash}}]{stochasticbank}%
  \BibitemOpen
  \bibfield  {author} {\bibinfo {author} {\bibfnamefont {I.~W.}\ \bibnamefont
  {Harry}}, \bibinfo {author} {\bibfnamefont {B.}~\bibnamefont {Allen}},\ and\
  \bibinfo {author} {\bibfnamefont {B.~S.}\ \bibnamefont {Sathyaprakash}},\
  }\href {https://doi.org/10.1103/PhysRevD.80.104014} {\bibfield  {journal}
  {\bibinfo  {journal} {Phys. Rev. D}\ }\textbf {\bibinfo {volume} {80}},\
  \bibinfo {pages} {104014} (\bibinfo {year} {2009})}\BibitemShut {NoStop}%
\bibitem [{\citenamefont {Roy}\ \emph {et~al.}(2017)\citenamefont {Roy},
  \citenamefont {Sengupta},\ and\ \citenamefont {Thakor}}]{soumen}%
  \BibitemOpen
  \bibfield  {author} {\bibinfo {author} {\bibfnamefont {S.}~\bibnamefont
  {Roy}}, \bibinfo {author} {\bibfnamefont {A.~S.}\ \bibnamefont {Sengupta}},\
  and\ \bibinfo {author} {\bibfnamefont {N.}~\bibnamefont {Thakor}},\ }\href
  {https://doi.org/10.1103/PhysRevD.95.104045} {\bibfield  {journal} {\bibinfo
  {journal} {Phys. Rev. D}\ }\textbf {\bibinfo {volume} {95}},\ \bibinfo
  {pages} {104045} (\bibinfo {year} {2017})}\BibitemShut {NoStop}%
\bibitem [{\citenamefont {Roy}\ \emph {et~al.}(2019)\citenamefont {Roy},
  \citenamefont {Sengupta},\ and\ \citenamefont {Ajith}}]{soumen2}%
  \BibitemOpen
  \bibfield  {author} {\bibinfo {author} {\bibfnamefont {S.}~\bibnamefont
  {Roy}}, \bibinfo {author} {\bibfnamefont {A.~S.}\ \bibnamefont {Sengupta}},\
  and\ \bibinfo {author} {\bibfnamefont {P.}~\bibnamefont {Ajith}},\ }\href
  {https://journals.aps.org/prd/abstract/10.1103/PhysRevD.99.024048} {\bibfield
   {journal} {\bibinfo  {journal} {Phys. Rev. D}\ }\textbf {\bibinfo {volume}
  {99}},\ \bibinfo {pages} {024048} (\bibinfo {year} {2019})}\BibitemShut
  {NoStop}%
\bibitem [{\citenamefont {Harry}\ and\ \citenamefont
  {Fairhurst}(2011)}]{ianharrycwb}%
  \BibitemOpen
  \bibfield  {author} {\bibinfo {author} {\bibfnamefont {I.~W.}\ \bibnamefont
  {Harry}}\ and\ \bibinfo {author} {\bibfnamefont {S.}~\bibnamefont
  {Fairhurst}},\ }\href {https://doi.org/10.1103/PhysRevD.83.084002} {\bibfield
   {journal} {\bibinfo  {journal} {Phys. Rev. D}\ }\textbf {\bibinfo {volume}
  {83}},\ \bibinfo {pages} {084002} (\bibinfo {year} {2011})}\BibitemShut
  {NoStop}%
\bibitem [{\citenamefont {Keppel}(2013{\natexlab{a}})}]{Keppel1}%
  \BibitemOpen
  \bibfield  {author} {\bibinfo {author} {\bibfnamefont {D.}~\bibnamefont
  {Keppel}},\ }\href {https://doi.org/10.1103/PhysRevD.87.124003} {\bibfield
  {journal} {\bibinfo  {journal} {Phys. Rev. D}\ }\textbf {\bibinfo {volume}
  {87}},\ \bibinfo {pages} {124003} (\bibinfo {year}
  {2013}{\natexlab{a}})}\BibitemShut {NoStop}%
\bibitem [{\citenamefont {Keppel}(2013{\natexlab{b}})}]{Keppel2}%
  \BibitemOpen
  \bibfield  {author} {\bibinfo {author} {\bibfnamefont {D.}~\bibnamefont
  {Keppel}},\ }\href {https://arxiv.org/abs/1307.4158} {} (\bibinfo {year}
  {2013}{\natexlab{b}}),\ \Eprint {https://arxiv.org/abs/1307.4158}
  {arXiv:1307.4158} \BibitemShut {NoStop}%
\bibitem [{\citenamefont {Buonanno}\ \emph {et~al.}(2003)\citenamefont
  {Buonanno}, \citenamefont {Chen},\ and\ \citenamefont
  {Vallisneri}}]{fittingfactor}%
  \BibitemOpen
  \bibfield  {author} {\bibinfo {author} {\bibfnamefont {A.}~\bibnamefont
  {Buonanno}}, \bibinfo {author} {\bibfnamefont {Y.}~\bibnamefont {Chen}},\
  and\ \bibinfo {author} {\bibfnamefont {M.}~\bibnamefont {Vallisneri}},\
  }\href {https://doi.org/10.1103/PhysRevD.67.104025} {\bibfield  {journal}
  {\bibinfo  {journal} {Phys. Rev. D}\ }\textbf {\bibinfo {volume} {67}},\
  \bibinfo {pages} {104025} (\bibinfo {year} {2003})}\BibitemShut {NoStop}%
\bibitem [{\citenamefont {Kalaghatgi}\ \emph {et~al.}(2015)\citenamefont
  {Kalaghatgi}, \citenamefont {Ajith},\ and\ \citenamefont {Arun}}]{chinm}%
  \BibitemOpen
  \bibfield  {author} {\bibinfo {author} {\bibfnamefont {C.}~\bibnamefont
  {Kalaghatgi}}, \bibinfo {author} {\bibfnamefont {P.}~\bibnamefont {Ajith}},\
  and\ \bibinfo {author} {\bibfnamefont {K.~G.}\ \bibnamefont {Arun}},\ }\href
  {https://doi.org/10.1103/PhysRevD.91.124042} {\bibfield  {journal} {\bibinfo
  {journal} {Phys. Rev. D}\ }\textbf {\bibinfo {volume} {91}},\ \bibinfo
  {pages} {124042} (\bibinfo {year} {2015})}\BibitemShut {NoStop}%
\bibitem [{\citenamefont {Sengupta}\ \emph {et~al.}(2002)\citenamefont
  {Sengupta}, \citenamefont {Dhurandhar}, \citenamefont {Lazzarini},\ and\
  \citenamefont {Prince}}]{anandsen}%
  \BibitemOpen
  \bibfield  {author} {\bibinfo {author} {\bibfnamefont {A.~S.}\ \bibnamefont
  {Sengupta}}, \bibinfo {author} {\bibfnamefont {S.~V.}\ \bibnamefont
  {Dhurandhar}}, \bibinfo {author} {\bibfnamefont {A.}~\bibnamefont
  {Lazzarini}},\ and\ \bibinfo {author} {\bibfnamefont {T.}~\bibnamefont
  {Prince}},\ }\href {https://doi.org/10.1088/0264-9381/19/7/337} {\bibfield
  {journal} {\bibinfo  {journal} {Classical Quantum Gravity}\ }\textbf
  {\bibinfo {volume} {19}},\ \bibinfo {pages} {1507} (\bibinfo {year}
  {2002})}\BibitemShut {NoStop}%
\bibitem [{\citenamefont {Abbott}\ \emph
  {et~al.}(2016{\natexlab{d}})\citenamefont {Abbott} \emph
  {et~al.}}]{detchar1}%
  \BibitemOpen
  \bibfield  {author} {\bibinfo {author} {\bibfnamefont {B.~P.}\ \bibnamefont
  {Abbott}} \emph {et~al.},\ }\href
  {https://doi.org/10.1088/0264-9381/33/13/134001} {\bibfield  {journal}
  {\bibinfo  {journal} {Classical Quantum Gravity}\ }\textbf {\bibinfo {volume}
  {33}},\ \bibinfo {pages} {134001} (\bibinfo {year}
  {2016}{\natexlab{d}})}\BibitemShut {NoStop}%
\bibitem [{\citenamefont {Zevin}\ \emph {et~al.}(2017)\citenamefont {Zevin},
  \citenamefont {Coughlin}, \citenamefont {Bahaadini}, \citenamefont {Besler},
  \citenamefont {Rohani}, \citenamefont {Allen}, \citenamefont {Cabero},
  \citenamefont {Crowston}, \citenamefont {Katsaggelos}, \citenamefont {Larson}
  \emph {et~al.}}]{detchar2}%
  \BibitemOpen
  \bibfield  {author} {\bibinfo {author} {\bibfnamefont {M.}~\bibnamefont
  {Zevin}}, \bibinfo {author} {\bibfnamefont {S.}~\bibnamefont {Coughlin}},
  \bibinfo {author} {\bibfnamefont {S.}~\bibnamefont {Bahaadini}}, \bibinfo
  {author} {\bibfnamefont {E.}~\bibnamefont {Besler}}, \bibinfo {author}
  {\bibfnamefont {N.}~\bibnamefont {Rohani}}, \bibinfo {author} {\bibfnamefont
  {S.}~\bibnamefont {Allen}}, \bibinfo {author} {\bibfnamefont
  {M.}~\bibnamefont {Cabero}}, \bibinfo {author} {\bibfnamefont
  {K.}~\bibnamefont {Crowston}}, \bibinfo {author} {\bibfnamefont {A.~K.}\
  \bibnamefont {Katsaggelos}}, \bibinfo {author} {\bibfnamefont {S.~L.}\
  \bibnamefont {Larson}}, \emph {et~al.},\ }\href
  {https://doi.org/10.1088/1361-6382/aa5cea} {\bibfield  {journal} {\bibinfo
  {journal} {Classical Quantum Gravity}\ }\textbf {\bibinfo {volume} {34}},\
  \bibinfo {pages} {064003} (\bibinfo {year} {2017})}\BibitemShut {NoStop}%
\bibitem [{\citenamefont {Huang}\ \emph {et~al.}(2018)\citenamefont {Huang},
  \citenamefont {Middleton}, \citenamefont {Ng}, \citenamefont {Vitale},\ and\
  \citenamefont {Veitch}}]{detchar3}%
  \BibitemOpen
  \bibfield  {author} {\bibinfo {author} {\bibfnamefont {Y.}~\bibnamefont
  {Huang}}, \bibinfo {author} {\bibfnamefont {H.}~\bibnamefont {Middleton}},
  \bibinfo {author} {\bibfnamefont {K.~K.~Y.}\ \bibnamefont {Ng}}, \bibinfo
  {author} {\bibfnamefont {S.}~\bibnamefont {Vitale}},\ and\ \bibinfo {author}
  {\bibfnamefont {J.}~\bibnamefont {Veitch}},\ }\href
  {https://doi.org/10.1103/PhysRevD.98.123021} {\bibfield  {journal} {\bibinfo
  {journal} {Phys. Rev. D}\ }\textbf {\bibinfo {volume} {98}},\ \bibinfo
  {pages} {123021} (\bibinfo {year} {2018})}\BibitemShut {NoStop}%
\bibitem [{\citenamefont {Abbott}\ \emph {et~al.}(2018)\citenamefont {Abbott}
  \emph {et~al.}}]{dqpaper}%
  \BibitemOpen
  \bibfield  {author} {\bibinfo {author} {\bibfnamefont {B.~P.}\ \bibnamefont
  {Abbott}} \emph {et~al.},\ }\href {https://doi.org/10.1088/1361-6382/aaaafa}
  {\bibfield  {journal} {\bibinfo  {journal} {Classical Quantum Gravity}\
  }\textbf {\bibinfo {volume} {35}},\ \bibinfo {pages} {065010} (\bibinfo
  {year} {2018})}\BibitemShut {NoStop}%
\bibitem [{\citenamefont {Allen}\ \emph {et~al.}(1999)\citenamefont {Allen},
  \citenamefont {Blackburn}, \citenamefont {Brady}, \citenamefont {Creighton},
  \citenamefont {Creighton}, \citenamefont {Droz}, \citenamefont {Gillespie},
  \citenamefont {Hughes}, \citenamefont {Kawamura}, \citenamefont {Lyons} \emph
  {et~al.}}]{allen}%
  \BibitemOpen
  \bibfield  {author} {\bibinfo {author} {\bibfnamefont {B.}~\bibnamefont
  {Allen}}, \bibinfo {author} {\bibfnamefont {J.~K.}\ \bibnamefont
  {Blackburn}}, \bibinfo {author} {\bibfnamefont {P.~R.}\ \bibnamefont
  {Brady}}, \bibinfo {author} {\bibfnamefont {J.~D.~E.}\ \bibnamefont
  {Creighton}}, \bibinfo {author} {\bibfnamefont {T.}~\bibnamefont
  {Creighton}}, \bibinfo {author} {\bibfnamefont {S.}~\bibnamefont {Droz}},
  \bibinfo {author} {\bibfnamefont {A.~D.}\ \bibnamefont {Gillespie}}, \bibinfo
  {author} {\bibfnamefont {S.~A.}\ \bibnamefont {Hughes}}, \bibinfo {author}
  {\bibfnamefont {S.}~\bibnamefont {Kawamura}}, \bibinfo {author}
  {\bibfnamefont {T.~T.}\ \bibnamefont {Lyons}}, \emph {et~al.},\ }\href
  {https://doi.org/10.1103/PhysRevLett.83.1498} {\bibfield  {journal} {\bibinfo
   {journal} {Phys. Rev. Lett.}\ }\textbf {\bibinfo {volume} {83}},\ \bibinfo
  {pages} {1498} (\bibinfo {year} {1999})}\BibitemShut {NoStop}%
\bibitem [{\citenamefont {Nitz}(2018)}]{nitzsgveto}%
  \BibitemOpen
  \bibfield  {author} {\bibinfo {author} {\bibfnamefont {A.~H.}\ \bibnamefont
  {Nitz}},\ }\href {https://doi.org/10.1088/1361-6382/aaa13d} {\bibfield
  {journal} {\bibinfo  {journal} {Classical Quantum Gravity}\ }\textbf
  {\bibinfo {volume} {35}},\ \bibinfo {pages} {035016} (\bibinfo {year}
  {2018})}\BibitemShut {NoStop}%
\bibitem [{\citenamefont {Nitz}\ \emph {et~al.}(2017)\citenamefont {Nitz},
  \citenamefont {Dent}, \citenamefont {Canton}, \citenamefont {Fairhurst},\
  and\ \citenamefont {Brown}}]{nitzphasetd}%
  \BibitemOpen
  \bibfield  {author} {\bibinfo {author} {\bibfnamefont {A.~H.}\ \bibnamefont
  {Nitz}}, \bibinfo {author} {\bibfnamefont {T.}~\bibnamefont {Dent}}, \bibinfo
  {author} {\bibfnamefont {T.~D.}\ \bibnamefont {Canton}}, \bibinfo {author}
  {\bibfnamefont {S.}~\bibnamefont {Fairhurst}},\ and\ \bibinfo {author}
  {\bibfnamefont {D.~A.}\ \bibnamefont {Brown}},\ }\href
  {https://doi.org/10.3847/1538-4357/aa8f50} {\bibfield  {journal} {\bibinfo
  {journal} {Astrophys. J.}\ }\textbf {\bibinfo {volume} {849}},\ \bibinfo
  {pages} {118} (\bibinfo {year} {2017})}\BibitemShut {NoStop}%
\bibitem [{\citenamefont {Biswas}\ \emph {et~al.}(2012)\citenamefont {Biswas},
  \citenamefont {Brady}, \citenamefont {Burguet-Castell}, \citenamefont
  {Cannon}, \citenamefont {Clayton}, \citenamefont {Dietz}, \citenamefont
  {Fotopoulos}, \citenamefont {Goggin}, \citenamefont {Keppel}, \citenamefont
  {Pankow}, \citenamefont {Price},\ and\ \citenamefont
  {Vaulin}}]{detectionstatistic}%
  \BibitemOpen
  \bibfield  {author} {\bibinfo {author} {\bibfnamefont {R.}~\bibnamefont
  {Biswas}}, \bibinfo {author} {\bibfnamefont {P.~R.}\ \bibnamefont {Brady}},
  \bibinfo {author} {\bibfnamefont {J.}~\bibnamefont {Burguet-Castell}},
  \bibinfo {author} {\bibfnamefont {K.}~\bibnamefont {Cannon}}, \bibinfo
  {author} {\bibfnamefont {J.}~\bibnamefont {Clayton}}, \bibinfo {author}
  {\bibfnamefont {A.}~\bibnamefont {Dietz}}, \bibinfo {author} {\bibfnamefont
  {N.}~\bibnamefont {Fotopoulos}}, \bibinfo {author} {\bibfnamefont {L.~M.}\
  \bibnamefont {Goggin}}, \bibinfo {author} {\bibfnamefont {D.}~\bibnamefont
  {Keppel}}, \bibinfo {author} {\bibfnamefont {C.}~\bibnamefont {Pankow}},
  \bibinfo {author} {\bibfnamefont {L.~R.}\ \bibnamefont {Price}},\ and\
  \bibinfo {author} {\bibfnamefont {R.}~\bibnamefont {Vaulin}},\ }\href
  {https://doi.org/10.1103/PhysRevD.85.122008} {\bibfield  {journal} {\bibinfo
  {journal} {Phys. Rev. D}\ }\textbf {\bibinfo {volume} {85}},\ \bibinfo
  {pages} {122008} (\bibinfo {year} {2012})}\BibitemShut {NoStop}%
\bibitem [{kan(2021)}]{kanchan}%
  \BibitemOpen
  \href@noop {} {\bibinfo {title} {{Hierarchical search pipeline}}},\ \bibinfo
  {howpublished}
  {\url{https://github.com/Kanchan-05/pycbc/tree/Hierarchical-search/hierarchical_search}}
  (\bibinfo {year} {2021})\BibitemShut {NoStop}%
\bibitem [{\citenamefont {Tiwari}(2018)}]{VTpycbc}%
  \BibitemOpen
  \bibfield  {author} {\bibinfo {author} {\bibfnamefont {V.}~\bibnamefont
  {Tiwari}},\ }\href {https://doi.org/10.1088/1361-6382/aac89d} {\bibfield
  {journal} {\bibinfo  {journal} {Classical Quantum Gravity}\ }\textbf
  {\bibinfo {volume} {35}},\ \bibinfo {pages} {145009} (\bibinfo {year}
  {2018})}\BibitemShut {NoStop}%
\bibitem [{\citenamefont {Sturani}\ \emph {et~al.}(2010)\citenamefont
  {Sturani}, \citenamefont {Fischetti}, \citenamefont {Cadonati}, \citenamefont
  {Guidi}, \citenamefont {Healy}, \citenamefont {Shoemaker},\ and\
  \citenamefont {Vicer{\'{e}}}}]{spintaylor}%
  \BibitemOpen
  \bibfield  {author} {\bibinfo {author} {\bibfnamefont {R.}~\bibnamefont
  {Sturani}}, \bibinfo {author} {\bibfnamefont {S.}~\bibnamefont {Fischetti}},
  \bibinfo {author} {\bibfnamefont {L.}~\bibnamefont {Cadonati}}, \bibinfo
  {author} {\bibfnamefont {G.~M.}\ \bibnamefont {Guidi}}, \bibinfo {author}
  {\bibfnamefont {J.}~\bibnamefont {Healy}}, \bibinfo {author} {\bibfnamefont
  {D.}~\bibnamefont {Shoemaker}},\ and\ \bibinfo {author} {\bibfnamefont
  {A.}~\bibnamefont {Vicer{\'{e}}}},\ }\href
  {https://doi.org/10.1088/1742-6596/243/1/012007} {\bibfield  {journal}
  {\bibinfo  {journal} {Journal of Physics: Conference Series}\ }\textbf
  {\bibinfo {volume} {243}},\ \bibinfo {pages} {012007} (\bibinfo {year}
  {2010})}\BibitemShut {NoStop}%
\bibitem [{lal(2018)}]{lalsuite}%
  \BibitemOpen
  \href {https://doi.org/https://doi.org/10.7935/GT1W-FZ16} {\bibinfo {title}
  {Ligo scientific collaboration, ligo algorithm library— lalsuite, free
  software (gpl)}} (\bibinfo {year} {2018})\BibitemShut {NoStop}%
\bibitem [{\citenamefont {Walt}\ \emph {et~al.}(2011)\citenamefont {Walt},
  \citenamefont {Colbert},\ and\ \citenamefont {Varoquaux}}]{numpy}%
  \BibitemOpen
  \bibfield  {author} {\bibinfo {author} {\bibfnamefont {S.~v.~d.}\
  \bibnamefont {Walt}}, \bibinfo {author} {\bibfnamefont {S.~C.}\ \bibnamefont
  {Colbert}},\ and\ \bibinfo {author} {\bibfnamefont {G.}~\bibnamefont
  {Varoquaux}},\ }\href {https://ieeexplore.ieee.org/document/5725236/}
  {\bibfield  {journal} {\bibinfo  {journal} {Comput. Sci. Eng.}\ }\textbf
  {\bibinfo {volume} {13}} (\bibinfo {year} {2011})}\BibitemShut {NoStop}%
\bibitem [{\citenamefont {Virtanen}\ \emph {et~al.}(2020)\citenamefont
  {Virtanen} \emph {et~al.}}]{scipy}%
  \BibitemOpen
  \bibfield  {author} {\bibinfo {author} {\bibfnamefont {P.}~\bibnamefont
  {Virtanen}} \emph {et~al.},\ }\href
  {http://www.nature.com/articles/s41592-019-0686-2} {\bibfield  {journal}
  {\bibinfo  {journal} {Nat. Methods}\ }\textbf {\bibinfo {volume} {17}}
  (\bibinfo {year} {2020})}\BibitemShut {NoStop}%
\bibitem [{\citenamefont {Price-Whelan}\ \emph {et~al.}(2018)\citenamefont
  {Price-Whelan} \emph {et~al.}}]{astropy}%
  \BibitemOpen
  \bibfield  {author} {\bibinfo {author} {\bibfnamefont {A.~M.}\ \bibnamefont
  {Price-Whelan}} \emph {et~al.},\ }\href
  {https://iopscience.iop.org/article/10.3847/1538-3881/aabc4f} {\bibfield
  {journal} {\bibinfo  {journal} {Astron. J.}\ }\textbf {\bibinfo {volume}
  {156}} (\bibinfo {year} {2018})}\BibitemShut {NoStop}%
\end{thebibliography}%

\end{document}